\documentstyle[aps,prbbib,twocolumn,epsf]{revtex}

\begin{document}
\draft
\title{Spin polarized parametric pumping}

\author{Wu Junling, Baigeng Wang, and Jian Wang$^{a)}$}
\address{Department of Physics, The University of Hong Kong, 
Pokfulam Road, Hong Kong, China\\
}
\maketitle

\begin{abstract}
We have developed a general theory for a parametric pump consisting of 
a nonmagnetic system with two ferromagnetic leads whose magnetic moments 
orient at an angle $\theta$ with respect to each other. In this
theory, the leads can maintain at different chemical potential. As a
result, the current is driven due to both the external bias and the
pumping potentials. When both $\theta$ and external bias are zero, our
theory recovers the known theory. In particular, two cases are 
considered: (a). in the adiabatic regime, we have derived the
pumped current for arbitrary pumping amplitude and external bias. (b).
at finite frequency, the system is away from equilibrium, we have
derived the pumped current up to quadratic order in pumping amplitude.  
Numerical results show interesting spin valve effects for pumped
current. 
\end{abstract}


\section{Introduction}
Recently, there is considerable interest in the parametric 
pumping\cite{brouwer,aleiner,switkes,zhou,shutenko,wei1,aleiner1,avron1,levinson,brouwer2,buttiker1,avron2,wang1,vavilov,wei2,renzoni,makhlin,chu,wbg1,wang2,levinson2,levinson1,buttiker2,vavilov1,entin1,wbg3,wbg2,blaauboer}. 
The parametric pump is facilitated by cyclic variations of pumping 
potentials inside the scattering system and has been realized
experimentally by Switkes et al\cite{switkes}. On theoretical side,
much progress has been made towards understanding of various features
related to the parametric pump. This includes quantization of pumped 
charge\cite{aleiner,levinson,wang2,entin1}, the influence of discrete 
spatial symmetries and magnetic field\cite{shutenko,aleiner1}, the 
rectification of displacement current\cite{brouwer2}, as well as the 
inelastic scattering\cite{buttiker1} to the pumped current. 
The concept of optimal pump has been proposed with the lower bound for
the dissipation derived\cite{avron2}. Within the formalism of
time-dependent scattering matrix theory, the heat current and shot 
noise in the pumping process\cite{makhlin,buttiker2,vavilov1,wbg2} 
has also been discussed. Recently, the original adiabatic pumping theory 
has been extended to account for the effect due to finite 
frequency\cite{vavilov,wbg1}, Andreev reflection in the presence of 
superconducting lead\cite{wang1,blaauboer}, 
and strong electron interaction in the Kondo regime\cite{wbg3}. This
gives us more physical insight of parametric pumping. For instance, the
experimental observed anomaly of pumped current at $\phi=0$ and $\phi=\pi$
can be explained using finite frequency theory\cite{wbg1} as due to the 
quantum interference of different photon assisted processes. When
superconducting lead is present, the interference between the direct 
reflection and multiple Andreev reflection gives rise to an enhancement
of pumped current which is four times of that of normal system\cite{wang1}. 
It will be interesting to further extend the parametric theory to the
case where the ferromagnetic leads are present. With the theory
extended, many new physics are foreseen\cite{wu} which may lead to new 
operational paradigms for future spintronic devices\cite{review}. In 
this paper, we
have developed a parametric pumping theory for a nonmagnetic system with
two ferromagnetic leads whose magnetic moments orient at an angle 
$\theta$ with respect to each other. Our theory is based on nonequilibrium 
Green's function approach and focused on current perpendicular to plane 
geometry. Parametric pump generates current at zero external bias. It
would be interesting to see the interplay of the role played by pumping
potential and external bias if the leads are maintain at different
chemical potential\cite{levinson1}. Hence in our theory, the external
bias is also included. In the adiabatic regime, the pumped current is 
proportional to pumping frequency. In this regime, we have derived the 
parametric pumping theory for finite pumping amplitude. At the finite
pumping frequency, the system is away from equilibrium, we have performed
perturbation up to the second order in pumping amplitude and obtained 
the pumped current at finite frequencies. The newly developed theory
allows us to study the pumped current for a variety of parameters, such
as the pumping amplitude, pumping frequency, phase difference between
two pumping potentials, the angular dependence between the magnetization of
two leads, as well as the external bias. We have applied our theory 
to a tunneling magnetoresistance (TMR) junction\cite{slon}. Due to the 
reported room temperature operation of TMR, the fundmental principle and 
transport properties of TMR devices has attracted increasingly
attention\cite{moodera}. Our numerical results show interesting spin valve
effect for pumped current. The paper is organized as
follows. In section II, we derive the general theory of a parametric
pump in the presence of ferromagnetic leads. The numerical results and 
summary are presented in section III.

\section{General theory}

The system we examine consists of a nonmagnetic system connected by two 
ferromagnetic 
electrodes to the reservoir. The magnetic moment ${\bf M}$ of the 
left electrode is pointing to the $z$-direction, the electric current 
is flowing in the $y$-direction, while the moment of the right 
electrode is at an angle $\theta$ to the $z$-axis in the $x-z$ plane. 
The Hamiltonian of the system is of the following form
\begin{equation}
H = H_L+H_R+H_0+V_p+H_T
\label{eqq1}
\end{equation}
where $H_L$ and $H_R$ describe the left and right electrodes 
\begin{equation}
H_L = \sum_{k\sigma} (\epsilon_{k L}+\sigma M) ~ c^{\dagger}_{kL\sigma}
c_{kL\sigma}
\label{eqq2}
\end{equation}
\begin{eqnarray}
H_R &=& \sum_{k\sigma} [(\epsilon_{k R}+\sigma M\cos\theta) ~
c^{\dagger}_{kR\sigma} c_{kR\sigma}  \nonumber \\
&+& \sum_{k\sigma} M\sin\theta ~ [c^{\dagger}_{kR\sigma} c_{kR\bar{\sigma}} ]
\ \ .
\label{eqq3}
\end{eqnarray}
In Eq. (\ref{eqq1}), $H_0$ describes the nonmagnetic (NM) scattering
region, 
\begin{equation}
H_0= \sum_{n\sigma} \epsilon_n d^{\dagger}_{n\sigma} d_{n\sigma}\ \ .
\label{hdot}
\end{equation}
$V_p$ is the time-dependent pumping potential and $H_T$ describes 
the coupling between electrodes and the NM scattering region 
with hopping matrix $T_{k\alpha n}$. To simplify the analysis, we 
assume the hopping matrix to be independent of spin index, hence 
\begin{equation}
H_T= \sum_{k \alpha n\sigma} [T_{k\alpha n} ~ 
c^{\dagger}_{k\alpha\sigma} d_{n\sigma} + c.c.]\ \ .
\label{ht}
\end{equation}
In these expressions $\epsilon_{k\alpha} = \epsilon^0_k + qV_\alpha$
with $\alpha = L,R$; $c^{\dagger}_{k\alpha\sigma}$ (with $\sigma=\uparrow,
\downarrow$ or $\pm 1$ and $\bar{\sigma}=-\sigma$) is the creation operator 
of electrons with spin index $\sigma$ inside the $\alpha$-electrode. 
Similarly $d^{\dagger}_{n\sigma}$ is the creation operator of electrons 
with spin $\sigma$ at energy level $n$ for the NM scattering region. In 
writing down Eqs.(\ref{eqq2}) and (\ref{eqq3}), we have made a 
simplification
that the value of molecular field $M$ is the same for the two electrodes, 
thus the spin-valve effect is obtained\cite{slon} by varying the angle 
$\theta$. Essentially, $M$ mimics the difference of density of states 
(DOS) between spin up and down electrons\cite{slon} in the electrodes. 
In this paper, we only consider the single electron behavior.
The charge quantization is not considered so that our system
is not in the Coulomb blockade regime. In addition, for the nonmagnetic
regions we are interested in, the Kondo effect can be neglected.

To proceed, we first apply the following Bogliubov transformation\cite{bog} 
to diagonalize the Hamiltonian of the right electrode\cite{wbg4}, 
\begin{equation}
c_{kR\sigma }\rightarrow \cos (\theta/2)C_{kR\sigma }-\sigma \sin (
\theta/2)C_{kR\bar{\sigma}}
\end{equation}
\begin{equation}
c_{kR\sigma }^{+}\rightarrow \cos (\theta/2)C_{kR\sigma
}^{+}-\sigma \sin (\theta/2)C_{kR\bar{\sigma}}^{+}
\end{equation}
from which we obtain the effective Hamiltonian
\begin{equation}
H_\alpha = \sum_{k\sigma} [(\epsilon_{k\alpha}+\sigma M)
C^{\dagger}_{k\alpha\sigma}C_{k\alpha\sigma}
\label{halpha}
\end{equation}
\begin{eqnarray}
H_T &=& \sum_{k n\sigma} [T_{kLn} ~ C^{\dagger}_{kL\sigma} 
d_{n\sigma} + T_{kRn} (\cos\frac{\theta}{2} ~ C^{\dagger}_{kR\sigma}  
\nonumber \\
&-& \sigma \sin\frac{\theta}{2} ~ C^{\dagger}_{kR\bar{\sigma}}) ~ 
d_{n\sigma} +c.c.]\ \ .
\label{eff}
\end{eqnarray}
In the following subsections, we will consider two cases: (1). parametric
pumping in the low frequency limit with finite pumping amplitude.
(2). pumping in the weak pumping limit with finite pumping frequency.

\subsection{Pumping in the low frequency limit}

In the subsection, we examine the pumping current at low frequency
limit while maintaining pumping amplitude finite. In this limit,
the system is nearly in equilibrium and we will use the equilibrium
Green's function\cite{datta,jauho,wbg5} to characterize the pumping process.
Using the distribution function, the total charge in the system during the
pumping is given by 
\begin{equation}
Q(x,t)=-iq\int (dE/2\pi )({\bf G}^{<}(E,\{V(t)\}))_{xx}
\end{equation}
where ${\bf G}^{<}$ is the lesser Green's function in real space, $x$
labels the position, and $\{V(t)\}$ describes a set of external parameters
which facilitates the pumping process. Within Hartree approximation, $
{\bf G}^{<}$ is related to the retarded and advanced Green's
functions ${\bf G}^{r}$ and ${\bf G}^{a}$, 
\begin{equation}
{\bf G}^{<}(E,\{V\})={\bf G}^{r}(E,\{V\})i\sum_\alpha 
{\bf \Gamma}_\alpha f_\alpha(E){\bf G}^{a}(E,\{V\})\ \ .
\end{equation}
where $f_\alpha(E) = f(E-qV_\alpha)$. In the low frequency limit, the 
retarded Green's function in real space is given by

\begin{equation}
{\bf G}^{r}(E,\{X\})=\frac{1}{E-H_0-{\bf V}_p-{\bf \Sigma}^{r}}
\end{equation}%
where ${\bf \Sigma}^{r}\equiv \sum_{\alpha} {\bf \Sigma}_{\alpha }^{r}$ 
is the self energy, and ${\bf \Gamma}_{\alpha}=
-2Im[{\bf \Sigma}_{\alpha}^{r}]$ is the linewidth function. 
In above equations, ${\bf G}^{r,a,<}$ denotes a $2\times 2$ matrix
with matrix elements $G_{\sigma, \sigma'}^{r,a<}$ and $\sigma=
\uparrow, \uparrow$. ${\bf V}_p = V_p {\bf I}$ where ${\bf I}$ is a
$2\times2$ unit matrix in the spin space.
In real space representation, $V_{p}$ is a diagonal matrix describing 
the variation of the potential landscape due to the external pumping 
parameter $V$. The self-energies are given\cite{wbg4}

\begin{equation}
{\bf \Sigma}^r_\alpha(E) = \hat{R}_\alpha \left( 
\begin{array}{cc}
\Sigma^r_{\alpha\uparrow} & 0 \\ 
0 & \Sigma^r_{\alpha\downarrow}
\end{array}
\right) \hat{R}^\dagger_\alpha
\label{self}
\end{equation}
and 
\begin{equation}
{\bf \Sigma}^<_\alpha(E) = i f_\alpha \hat{R}_\alpha \left( 
\begin{array}{cc}
\Gamma^0_{\alpha\uparrow} & 0 \\ 
0 & \Gamma^0_{\alpha\downarrow}
\end{array}
\right) \hat{R}_\alpha^\dagger
\label{ss}
\end{equation}
with the rotational matrix $\hat{R}_\alpha$ for electrode $\alpha$ 
defined as 
\begin{equation}
\hat{R} = \left( 
\begin{array}{cc}
\cos\theta_\alpha/2 ~~ & \sin\theta_\alpha/2 \\ 
-\sin\theta_\alpha/2 ~~ & \cos\theta_\alpha/2
\end{array}
\right)\ \ .
\end{equation}
Here angle $\theta_{\alpha}$ is defined as $\theta_L=0$ and 
$\theta_R=\theta$ and $\Sigma^r_{\alpha \sigma}$ is given by 
\begin{equation}
\Sigma^r_{\alpha \sigma mn}= \sum_k \frac{T^*_{k\alpha m} T_{k\alpha n}} 
{E-\epsilon^0_{k \alpha \sigma} +i\delta}
\end{equation}
and $\Gamma^0_{\alpha\sigma}= -2Im(\Sigma^r_{\alpha\sigma})$ is the
linewidth function when $\theta=0$. 

In order for a parametric electron pump to function at low
frequency, we need simultaneous variation of two or more 
system parameters controlled by gate voltages: $V_{i}(t)=V_{i0}+
V_{ip}\cos (\omega t +\phi_i)$. 
Hence, in our case, the potential due to the gates can be written as 
$V_{p}=\sum_i V_{i}{\bf \Delta }_{i}$, where ${\bf \Delta }_{i}$ 
is potential profile due to each pumping potential. For simplicity we 
assume a constant gate potential
such as that $({\bf \Delta }_{1})_{xx}$ is one for $x$ in the first gate
region and zero otherwise. If the time variation of these parameters are
slow, i.e. for $V(t)=V_{0}+\delta V\cos (\omega t)$, then the
charge of the system coming from all contacts due to the infinitesimal
change of the system parameter ($\delta V\rightarrow 0$) is

\begin{equation}
dQ_p(t)=\sum_{i}\partial_{V_{i}}Tr[Q(x,t)]~\delta V_{i}(t)
\end{equation}
it is easily seen that the total charge in the system in a period is zero
which is required for the charge conservation. To calculate the pumped
current, we have to find the charge $dQ_{p\alpha }$ passing through contact 
$\alpha $ due to the change of the system parameters. Using 
the Dyson equation $\partial_{V_{i}}{\bf G}^{r}={\bf G}^{r}\Delta _{i}
{\bf G}^{r}$, the above equation becomes,

\begin{eqnarray}
&& dQ_p(t) = q\sum_j \int \frac{dE}{2\pi} \sum_\beta {\rm Tr}[ {\bf G}^r 
{\bf \Delta}_j {\bf G}^r {\bf \Gamma}_\beta {\bf G}^a
\nonumber \\
&&+{\bf G}^r {\bf \Gamma}_\beta {\bf G}^a {\bf \Delta}_j {\bf G}^a 
] f_\beta(E) \delta V_j(t)
\nonumber \\
&&=-q\int \frac{dE}{2\pi} \sum_j \sum_\beta f_\beta
{\rm Tr} [ \partial_E[{\bf G}^r {\bf \Gamma}_\beta
{\bf G}^a {\bf \Delta}_j] ] \delta V_j(t) \nonumber
\end{eqnarray}
where the wideband limit has been taken. Integrating by part, we
obtain
\begin{equation}
dQ_p(t)=q\int dE \sum_\beta (\partial_E f_\beta) 
\sum_j \frac{dN_\beta}{dV_j} \delta V_j(t)
\end{equation}
where we have used the injectivity\cite{but5}
\begin{equation}
\frac{dN_\beta}{dV_j} = \frac{1}{2\pi} {\rm Tr} 
[{\bf G}^r {\bf \Gamma}_\beta {\bf G}^a {\bf \Delta}_j] 
\label{dndx}
\end{equation} 
Using the partial density of states $dN_{\alpha \beta}/dV_j$
defined as\cite{but6}

\begin{eqnarray}
&&\frac{dN_{\alpha \beta}}{dV_j} = \frac{1}{4\pi} {\rm Tr} [
{\bf G}^r {\bf \Gamma}_\alpha {\bf G}^r \Delta_j +c.c.]
\delta_{\alpha \beta} \nonumber \\
&&+\frac{1}{4\pi} {\rm Tr} [i {\bf G}^r {\bf \Gamma}_\beta
{\bf G}^a {\bf \Gamma}_\alpha {\bf G}^r \Delta_j +c.c.]
\end{eqnarray}
with $\sum_\alpha dN_{\alpha \beta}/dV_j = dN_{\beta}/dV_j$,
we obtain 

\begin{equation}
dQ_{p\alpha}(t)=-q\int dE \sum_\beta (-\partial_E f_\beta) 
\sum_j \frac{dN_{\alpha \beta}}{dV_j} \delta V_j(t)
\end{equation}
If we include the charge passing through contact $\alpha$ due the
external bias, then

\begin{eqnarray}
&&dQ_{\alpha}(t)=-q\int dE \sum_\beta (-\partial_E f_\beta) 
\sum_j \frac{dN_{\alpha \beta}}{dV_j} \frac{dV_j(t)}{dt} dt
\nonumber \\
&&-q\int dE \sum_\beta {\rm Tr} [{\bf \Gamma}_\alpha {\bf G}^r
{\bf \Gamma}_\beta {\bf G}^a] (f_\alpha - f_\beta) dt
\end{eqnarray}
Furthermore, the total current flowing through contact $\alpha$ due to 
both the variation of parameters $V_j$ and external bias, 
in one period, is given by
\begin{equation}
J_\alpha = \frac{1}{\tau} \int_0^{\tau} dt ~ dQ_\alpha/dt
\label{current}
\end{equation}
where $\tau=2\pi/\omega$ is the period of cyclic variation. If there
are two pumping parameters, Eq.(\ref{current}) can be written as
when $\alpha=1$,

\begin{eqnarray}
&&J^{(1)}_1 =\frac{q}{2\tau} \int dt \int dE ~ \partial_E (f_1-f_2)
\sum_j \nonumber \\
&&[ \frac{dN_{11}}{dV_j} \frac{dV_j(t)}{dt} - 
\frac{dN_{12}}{dV_j} \frac{dV_j(t)}{dt} ] dt \nonumber \\
&&-\frac{q}{\tau}\int dt \int dE {\rm Tr} [{\bf \Gamma}_1 {\bf G}^r
{\bf \Gamma}_2 {\bf G}^a] (f_1 - f_2) 
\label{cur1}
\end{eqnarray}
and 
\begin{eqnarray}
&&J^{(2)}_1 =\frac{q}{2\tau} \int dt \int dE ~ \partial_E (f_1+f_2)
\sum_j \nonumber \\
&&[ \frac{dN_{11}}{dV_j} \frac{dV_j(t)}{dt} + 
\frac{dN_{12}}{dV_j} \frac{dV_j(t)}{dt} ] dt 
\label{cur2}
\end{eqnarray}
where $J_1=J^{(1)}_1+J^{(2)}_2$. In Ref.\onlinecite{levinson1},
$J^{(1)}_1$ has been identified as the current due to the
external bias and $J^{(2)}$ as pumping current. For $\theta=0,\pi$,
all the $2\times 2$ matrix are diagonal. In this case, it is easy to
show that Eq.(\ref{current}) agrees with the result of 
Ref.\onlinecite{levinson1}. If the external bias is zero, 
Eq.(\ref{current}) reduces to the familiar formula\cite{brouwer}
when there are two pumping potentials,

\begin{equation}
J_\alpha = \frac{q\omega}{2\pi} \int_0^{\tau} dt \left[\frac{dN_\alpha}
{dX_1} \frac{dX_1}{dt} + \frac{dN_\alpha}{dX_2} \frac{dX_2}{dt}\right]
\label{pump}
\end{equation}

\subsection{Finite frequency pumping in the weak pumping limit}

In this subsection, we will calculate the pumping current at finite
frequency. The Keldysh nonequilibrium Green's function approach used 
here is in the standard tight-binding representation\cite{datta}.
We could not use the momentum space version because the time dependent 
perturbation (pumping potential) inside the scattering region is position 
dependent. Hence it is most suitable to use a tight-binding real space 
technique. In contrast, in previous investigations\cite{pre} on 
photon-assisted processes the time-dependent potential is uniform 
throughout the dot and therefore a momentum space method is easier to 
apply. 

Assuming the time-dependent perturbations located at the different sites, 
$j=i_0, j_0$, and $k_0$ etc, with 
\begin{equation}
V_j(t)=V_j\cos (\omega t+\phi_j)
\end{equation}
When there is no interaction between electrons in the ideal leads $L$ 
and $R$, the standard nonequilibrium Green's function theory gives the 
following expression for the time dependent current\cite{jauho},

\begin{eqnarray}
J_{\alpha }(t)&=&-q\int_{-\infty }^{t}dt_{1} 
{\rm Tr}[{\bf G}^{r}(t,t_{1}){\bf
\Sigma }_{\alpha }^{<}(t_{1},t) \nonumber \\
&+&{\bf G}^{<}(t,t_{1}){\bf \Sigma }
_{\alpha }^{a}(t_{1},t)+c.c.]
\end{eqnarray}
and transmission coefficient
\begin{equation}
T(E) ={\rm Tr} [{\bf \Gamma}_{L}^{r}(E)){\bf G}^{r}(E)  {\bf 
\Gamma}_{R}^{r}(E)){\bf G}^{a}(E)] 
\end{equation}
where the scattering Green's functions and self-energy are
defined in the usual manner: 
\begin{equation}
{\bf G}_{ij \sigma \sigma'}^{r,a}(t_{1},t_{2})=\mp 
i\theta (\pm t_{1}\mp t_{2})\langle
\{d_{i,\sigma}(t_{1}),d_{j,\sigma'}^{+}(t_{2})\}\rangle
\end{equation}

\begin{equation}
{\bf G}_{ij,\sigma,\sigma'}^{<}(t_{1},t_{2})=
i\langle d_{j,\sigma'}^{+}(t_{2})d_{i,\sigma,}(t_{1})\rangle
\end{equation}

\begin{equation}
{\bf \Sigma }_{\alpha ij}^{r,a,<}(t_{1},t_{2})=\sum_{k}T_{\alpha
ki}^{\ast }T_{\alpha kj}{\bf g}_{\alpha }^{r,a,<}(t_{1},t_{2})
\end{equation}

The average current $J_L(t)$ from the left lead can be written as 
\begin{eqnarray}
<J_{L}(t)>&=&-\frac{q}{\tau}\int_0^\tau dt \int_{-\infty }^{t}dt_{1}
Tr[{\bf G}_{11}^{r}(t,t_{1}){\bf \Sigma}_{L}^{<}(t_{1},t) \nonumber \\
&+&{\bf G}_{11}^{<}(t,t_{1}){\bf \Sigma}
_{L}^{a}(t_{1},t)+c.c.]
\label{average}
\end{eqnarray}

In the absence of pumping, the retarded Green's function
is defined in terms of the Hamiltonian $H_{0}$, 
\begin{equation}
{\bf G}^{0r}(E)=\frac{1}{E-H_0-{\bf \Sigma }^{r}}
\end{equation}
and ${\bf G}^{0<}$ is related to the retarded and advanced 
Green's functions ${\bf G}^{0r}$ and ${\bf G}^{0a}$ by,

\begin{equation}
{\bf G}^{0<}(E)={\bf G}^{0r}(E){\bf \Sigma}^{<}(E){\bf G}^{0a}(E)
\end{equation}
Now we make use of the time-dependent pumping potentials $V_j(t)$ 
as the perturbations to calculate all kinds of Green's functions up to 
the second order, and corresponding average current.

First, we calculate the current corresponding to the term 
${\bf G}_{11}^{r}(t,t_{1}) {\bf \Sigma}_{L}^{<}(t_{1},t)$ in
Eq.(\ref{average}).  Dyson equation for ${\bf G}_{11}^{r}(t,t_{1})$ 
gives the second order contribution:
\begin{eqnarray}
&&{\bf G}_{11}^{(2)r}(t,t_{1}) =\sum_{jk} 
\int \int dxdy  {\bf G}_{1j}^{0r}(t-x) \nonumber \\
&&{\bf V}_{j}(x){\bf G}_{jk}^{0r}(x-y){\bf V}_{k}(y)
{\bf G}_{k 1}^{0r}(y-t_{1}) \nonumber \\
&& \equiv \sum_{jk} {\bf G}_{1j}^{0r}
{\bf V}_{j} {\bf G}_{jk}^{0r} {\bf V}_k
{\bf G}_{k 1}^{0r}
\label{g2}
\end{eqnarray}

Substituting Eq.(\ref{g2}) into Eq.(\ref{average}) and completing
the integration over time $x,y,t_1,t$, it is not difficult to calculate 
the average current $<J_{L1}>$ due to the first term in Eq.(\ref{average}) 
(see appendix for details),
\begin{eqnarray}
&& <J_{L1}(t)>=-\sum_{jk} \frac{qV_{j}V_{k}}{4}
\int \frac{dE}{2\pi }Tr\left[{\bf \Sigma }_{L}^{<}
{\bf G}_{1j}^{0r} \right. \nonumber\\
&&\left. [{\bf G}_{jk}^{0r}(E_{-} )e^{i\Delta_{kj}} 
+{\bf G}_{jk}^{0r}(E_{+})e^{-i\Delta_{kj}}] {\bf G}_{k 1}^{0r} \right]  
\label{eq1}
\end{eqnarray}
where $\Delta_{kj}=\phi_{k}-\phi_{j}$ is the phase difference,  
$E_{\pm}=E\pm\omega$, and $G^{r,a,<}\equiv G^{r,a,<}(E)$. 

Now we calculate the second term ${\bf G}_{11}^{<}(t,t_{1}) 
{\bf \Sigma }_{L}^{a}(t_{1},t)$ in Eq.(\ref{average}). Using Keldysh 
equation, ${\bf G}^<={\bf G}^r{\bf \Sigma}^< {\bf G}^a$ , we have 
\begin{equation}
{\bf G}_{jk}^{<} ={\bf G}_{j 1}^{r} {\bf \Sigma }_{L}^{<} 
{\bf G}_{1k}^{a}  
+{\bf G}_{j N}^{r} {\bf \Sigma }_{R}^{<} {\bf G}_{N k}^{a}
\label{less}
\end{equation}
where ${\bf \Sigma}_\alpha^<=i{\bf \Gamma}_\alpha f_\alpha$.
Expanding $G^{r,a}$ up to the second order in pumping parameters, 
we obtain the second order contribution from ${\bf G}^<_{11}$,

\begin{eqnarray}
&&{\bf G}_{11}^{(2)<}={\bf G}_{11}^{r} {\bf \Sigma }_{L}^{<} 
{\bf G}_{11}^a +{\bf G}_{1N}^r {\bf \Sigma}_{R}^< {\bf G}_{N1}^a 
\nonumber \\
&&= {\bf G}_{11}^{(2)r}  {\bf \Sigma }_{L}^{<} {\bf G}_{11}^{0a}
+{\bf G}_{11}^{(1)r} {\bf \Sigma }_{L}^{<} {\bf G}_{11}^{(1)a}
+{\bf G}_{11}^{0r} {\bf \Sigma }_{L}^{<} {\bf G}_{11}^{(2)a}
\nonumber \\
&&+{\bf G}_{1N}^{(2)r}{\bf \Sigma}_{R}^{<} {\bf G}_{N1}^{0a} 
+{\bf G}_{1N}^{(1)r} {\bf \Sigma }_{R}^{<} {\bf G}_{N1}^{(1)a} 
+{\bf G}_{1N}^{0r} {\bf \Sigma }_{R}^{<} {\bf G}_{N1}^{(2)a}
\nonumber \\
&=& \sum_{jk} [{\bf G}_{1j}^{0r}{\bf V}_{j}{\bf G}_{jk}^{0r}
{\bf V}_{k} {\bf G}_{k 1}^{0<} +{\bf G}_{1j}^{0r}{\bf V}_j
{\bf G}_{jk}^{0<} {\bf V}_k {\bf G}_{k 1}^{0a} \nonumber \\
&&+ {\bf G}_{1j}^{0<} {\bf V}_j {\bf G}_{jk}^{0a} 
{\bf V}_k {\bf G}_{k 1}^{0a}]
\label{g22}
\end{eqnarray}
where we have used Eq.(\ref{less}) to simplify the expression.
After some algebra, we have the following three expressions  
corresponding to each term in Eq.(\ref{g22}), 

\begin{eqnarray}
&&-\sum_{jk} \frac{qV_j V_k}{4} \int \frac{dE}{2\pi} {\rm Tr} 
\left[{\bf \Sigma }_{L}^{a} {\bf G}_{1j}^{0r} \right. \nonumber \\
&&\left. [{\bf G}_{jk}^{0r}(E_{-})e^{i\Delta_{kj}}
+e^{-i\Delta_{kj}}{\bf G}_{jk}^{0r}(E_+)] {\bf G}_{k 1}^{0<} \right] 
\label{eq2}
\end{eqnarray}

\begin{eqnarray}
&&-\sum_{jk} \frac{qV_j V_k}{4} \int \frac{dE}{2\pi }Tr \left[
{\bf \Sigma }_{L}^{a} {\bf G}_{1j}^{0r}  \right. \nonumber \\
&& \left. [ {\bf G}_{jk}^{0<}(E_{-}) e^{i\Delta_{kj}}
+e^{-i\Delta_{kj}} {\bf G}_{jk}^{0<}(E_+)] {\bf G}_{k 1}^{0a}
\right]
\label{eq3}
\end{eqnarray}

\begin{eqnarray}
&&-\sum_{jk} \frac{qV_j V_k}{4} \int \frac{dE}{2\pi }Tr \left[
{\bf \Sigma}_{L}^{a} {\bf G}_{1j}^{0<} \right. \nonumber \\
&& \left. [{\bf G}_{jk}^{0a}(E_{-}) e^{i\Delta_{kj}} 
+e^{-i\Delta_{kj}}{\bf G}_{jk}^{0a}(E_+)] {\bf G}_{k 1}^{0a} \right]
\label{eq4}
\end{eqnarray}
The final pumped current is the sum of Eq.(\ref{eq1}), (\ref{eq2}), 
(\ref{eq3}), (\ref{eq4}) and their complex conjugates, in addition to 
the current directly due to the external bias (see second term of
Eq.(\ref{cur1})). If the external bias is zero, the expression of the 
pumped current can be simplified significantly.
Note that in the equilibrium, the lesser Green's function satisfys
the fluctuation-dissipation theorem, 
\begin{equation}
{\bf G}^{0<}(E)=-f(E)\left[ {\bf G}^{0r}(E)-{\bf G}^{0a}(E)\right]
\end{equation}
and 
\begin{equation}
{\bf G}^{0<}(E_{\pm})=-f(E_{\pm})\left[ {\bf G}^{0r}(E_{\pm})
-{\bf G}^{0a}(E_{\pm})\right]
\end{equation}

(\ref{eq1})+(\ref{eq2})+(\ref{eq4})$^*$ leads to 

\begin{eqnarray}
&&-i\sum_{jk} \frac{qV_j V_k}{4} \int \frac{dE}{2\pi } {\rm Tr} 
\left[{\bf \Gamma}_L {\bf G}_{1j}^{0r} f(E) \right. \nonumber \\
&& \left. [{\bf G}_{jk}^{0r}(E_{-})e^{i\Delta_{kj}}
+e^{-i\Delta_{kj}}{\bf G}_{kj}^{0r}(E_+)] {\bf G}_{k 1}^{0a} \right] 
\label{eq5}
\end{eqnarray}

while (\ref{eq1})$^*$+(\ref{eq2})$^*$+(\ref{eq4}) gives to
\begin{eqnarray}
&&i\sum_{jk} \frac{qV_j V_k}{4} \int \frac{dE}{2\pi } {\rm Tr} 
\left[{\bf \Gamma}_L {\bf G}_{1j}^{0r} f(E) \right. \nonumber \\
&& \left. [{\bf G}_{jk}^{0a}(E_{-})e^{i\Delta_{kj}}
+e^{-i\Delta_{kj}}{\bf G}_{jk}^{0a}(E_+)] {\bf G}_{k 1}^{0a} \right] 
\label{eq6}
\end{eqnarray}

furthermore, (\ref{eq3})+c.c. becomes 
\begin{eqnarray}
&&i\sum_{jk} \frac{qV_j V_k}{4} \int \frac{dE}{2\pi} {\rm Tr} 
\left[{\bf \Gamma}_L {\bf G}_{1j}^{0r} \right. \nonumber \\
&& \left. [f_{-} ({\bf G}_{jk}^{0r}(E_{-})
-{\bf G}_{jk}^{0a}(E_{-})) e^{i\Delta_{kj}} \right. \nonumber \\
&&\left.+e^{-i\Delta_{kj}} f_+ ({\bf G}_{jk}^{0r}(E_+)- 
{\bf G}_{jk}^{0a}(E_+))] {\bf G}_{k 1}^{0a} \right] 
\label{eq7}
\end{eqnarray}
where $f_{\pm}=f(E_{\pm})$. 
 
Combining Eqs.(\ref{eq5}), (\ref{eq6}) and (\ref{eq7}), we finally
obtain
\begin{eqnarray}
&&J_L=i\sum_{jk} \frac{qV_j V_k}{4} \int \frac{dE}{2\pi} {\rm Tr} 
\left[{\bf \Gamma}_L {\bf G}_{1j}^{0r} \right. \nonumber \\
&& \left. [(f_{-}-f) ({\bf G}_{jk}^{0r}(E_{-})
-{\bf G}_{jk}^{0a}(E_{-})) e^{i\Delta_{kj}} \right. \nonumber \\
&&\left.+e^{-i\Delta_{kj}} (f_+-f) ({\bf G}_{jk}^{0r}(E_+)- 
{\bf G}_{jk}^{0a}(E_+))] {\bf G}_{k 1}^{0a} \right] 
\label{eq8}
\end{eqnarray}
In the limit of small frequency, we expand Eq.(\ref{eq8}) up to the
first order in frequency and use the fact that

\begin{equation}
{\bf G}^{0r}-{\bf G}^{0a}=-i{\bf G}^{0r} {\bf \Gamma} {\bf G}^{0a}
\end{equation}
and Dyson equation,
\begin{equation}
{\bf G}_{ji}^{0r} {\bf G}_{ik}^{0r}=\frac{\partial 
{\bf G}_{jk}^{0r}}{\partial {\bf V_{i}}}
\end{equation}
we obtain, 
\begin{eqnarray}
&&J_L=\sum_{jk} \frac{q\omega V_j V_k \sin (\Delta_{kj})}{2} 
\int \frac{dE}{2\pi } \partial_{E} f(E) \nonumber \\
&& {\rm Tr} \left\{ {\bf \Gamma }_{L} {\bf G}_{1j}^{0r}(E)
\left[ {\bf G}_{jk}^{0r}(E)-{\bf G}_{jk}^{0a}(E)
\right] {\bf G}_{k 1}^{0a}(E)\right\} \nonumber \\
&&=-\sum_{jk} \frac{iq\omega V_j V_k \sin (\Delta_{jk})}{2}
\int \frac{dE}{2\pi}\partial_{E}f(E) \nonumber \\
&&{\rm Tr} \left[{\bf \Gamma }_{L}\frac{\partial 
{\bf G}_{11}^{0r}}{\partial {\bf V}_j} {\bf \Gamma }_{L}
\frac{\partial {\bf G}_{11}^{0a}}{\partial {\bf V}_k}
+{\bf \Gamma }_{L} \frac{\partial {\bf G}_{12}^{0r}}
{\partial {\bf V}_j} {\bf \Gamma }_{R}
\frac{\partial {\bf G}_{21}^{0a}}{\partial {\bf V}_{k}}\right]
\end{eqnarray}
which is the same as Ref.\onlinecite{brouwer} when $\theta=0$. 

\section{Results}
We now apply our formula Eqs.(\ref{cur1}), (\ref{cur2}) and (\ref{eq8})
to a TMR junction. For current perpendicular
to the plane geometry, the TMR junction can be modeled by an 
one-dimensional quantum structure with a double barrier potential 
$U(x)=X_1 \delta (x+a)+X_2 \delta (x-a)$ where $2a$ is the well width. 
For this system the Green's function $G(x,x')$ can be calculated 
exactly\cite{yip}. The adiabatic pump that we consider 
is operated by changing barrier heights adiabatically and periodically: 
$X_1=V_0+V_p\sin(\omega t)$ and $X_2=V_0+V_p\sin(\omega t+\phi)$.  
In the following, we will study zero temperature behavior of the pumped 
current. In the calculation, we have chosen $M=37.0$ and $V_0=79.2$.
Finally the unit is set by $\hbar=2m=1$\cite{foot2}.

We first study the pumped current with two pumping potentials in the
adiabatic regime.  Fig.1 depicts the transmission coefficient $T$
versus Fermi energy at several angles $\theta$. As expected, among different
angles, $T$ is the largest at $\theta=0$ and smallest at $\theta=\pi$
with the ratio $T_{max}(0)/T_{max}(\pi) \sim 4$. This gives the usual 
spin valve effect\cite{slon}. Fig.2 plots the pumped current versus
Fermi energy at different $\theta$. Here we have set the phase difference
of two pumping potentials to be $\pi/2$. Similar to the transmission 
coefficient, we obtain largest pumped current at $\theta=0$ and smallest
current at $\theta=\pi$. We found that the ratio $I_{max}(0)/I_{max}(\pi)$
is about the same as that of transmission coefficient. As the pumping 
amplitude doubles, the peak of pumped current is broadened and the maximum
pumped current is doubled (see inset of Fig.2). This is understandable
since at large pumping amplitude the instantaneous resonant level 
oscillates with a large amplitude and hence can generate heat current 
in a broad range of energy. The spin valve effect of 
pumped current is illustrated in Fig.3 where the pumped current versus
$\theta$ is shown when the system is at resonance. We see that the pumped
current is maximum at $\theta=0$ and decreases quickly as one increases
$\theta$ from $0$ to $\pi$. For larger pumping amplitude, we have similar
behavior (inset of Fig.3). In Fig.4, we plot the pumped current as a 
function of phase difference $\phi$ between two pumping potentials. 
We see that the pumped current is antisymmetric about the $\phi=\pi$. 
The  nonlinear behavior is clearly seen which deviates from the 
sinusoidal behavior at small pumping amplitude. In Fig.5, we show the 
pumped current in the presence of external bias. In the calculation we 
assume that $V_L=-\omega V/2$ and $V_R=\omega V/2$ so that the external 
bias is against the pumped current when $V$ is positive. Due to the 
external bias, the total pumped current (dashed line) decreases near
resonant energy and 
reverses the direction at other energies. Now we turn to the case of finite
frequency pumping. We first present our results (Fig.6 to Fig.8) at 
small pumping frequency $\omega=0.002$. In Fig.6 we plot the pumped 
current as a function of phase difference $\phi$ near the resonant energy. 
At $\theta=0$, the magnitude of pumped current is much larger than 
that at $\theta=\pi/2$ or $\pi$. We notice that at $\phi=0$ 
and $\phi=\pi$, the pumped current is nonzero similar to the experimental 
anomaly observed experimentally for nonmagnetic system\cite{switkes}. The 
pumped current away from resonant energy is shown in Fig.7. At
$\theta=0$, we see that the pumped current is sharply peaked at resonant 
energy. The pumped current is positive for $\phi=\pi/2$ and negative 
for $\phi=0$ and $\pi$. At $\theta=\pi/2$ or $\pi$, the pumped current at
$\phi=\pi/2$ (dashed line) is much larger than that at other angles. 
Fig.8 displays the pumped current as a function of $\theta$ near resonant 
energy.  For $\phi=\pi/2$ (dotted line), we see the usual behavior 
that large pumped current 
occurs at $\theta=0$ and it decreases to the minimum at $\theta=\pi$. 
For $\phi=0$ or $\pi$, however, we see completely different behavior. 
The pumped current is the still the largest at $\theta=0$ but the
direction of the pumped current is reversed.  As one increases 
$\theta$, the pumped current decreases and reaches a flat region with
almost zero pumped current. Now 
we study the effect of frequency to the pumped current versus
$\theta$ (see Fig.9). We will fix the phase difference to be 
$\phi=\pi/2$ and energy near resonance. At small frequency $\omega=0.002$
(dashed line), the pumped current versus $\theta$ show usual behavior. 
When the frequency is increased to $\omega=0.004$ (dot-dashed line), two 
peaks show up symmetrically near $\theta=\pm \pi/4$ while the minimum 
is still at $\theta=\pi$. As frequency is increased further to 
$\omega=0.006$ (dotted line), the pumped current near 
$\theta=0$ reverses the direction and the new peak position shifts to 
$\theta \sim 0.35\pi$.  Upon further increasing $\omega$, the curve of 
pumped current versus $\theta$ develops a flat region between 
$\theta=\pm 0.35\pi$ with positive current while the magnitude of the 
negative pumped current at $\theta=0$ becomes larger (see Fig.9b). 
Finally, at even 
larger frequency $\omega=0.1$, all the pumped currents are negative. 
This behavior can be understood from the photon assisted process\cite{wbg1}. 
The quantum interference between contributions due to photon emission 
(or absorption) near two pumping potentials is essential to understand 
the nature of pumped current. In addition to the interference effect, 
the pumped current is also affected by a competition of between the photon 
emission and absorption processes which tend to cancel to each other. It 
is the interplay between this competition and interference that gives rise 
to the interesting spin valve effect for the pumped current.
Finally, we show in Fig.10 the pumped current versus pumping frequency. 
We see that at small frequency the current is positive and small, at large
frequency the current is much large and is negative. 

In summary, we have developed a general theory for parametric pumping in
the presence of ferromagnetic leads. Our theory is based on the
nonequilibrium Green's function method and is valid for multi-modes 
(in two or three dimensions) and can be easily extended to the case of 
multi-probes (although most of
calculations are for two probes). In the parametric pumping, 
two kinds of driving forces are present: multiple pumping potentials 
inside the scattering system as well as the external bias in the 
multi-probes. Two cases are considered. In the adiabatic regime, the
system is in near equilibrium.  In this case our theory is for general 
pumping amplitude. At finite frequency, the system is away from
equilibrium. Our theory is up the quadratic order in pumping amplitude.
This theory allows us to examine the pumped current in broader
parameter space including pumping amplitude, pumping frequency, phase
difference between two pumping potentials, the angle between
magnetization of two leads.

\section*{Acknowledgments}
We gratefully acknowledge support by a RGC grant from the SAR Government of 
Hong Kong under grant number HKU 7091/01P.

\section{Appendix}

Now we show that 
\begin{eqnarray}
&&B\equiv \frac{1}{\tau}\int_0^\tau dt
\int_{-\infty }^{t}dt_{1} {\rm Tr} [{\bf F}_0(t_1,t)
{\bf G}(t,t_1)]
\nonumber \\
&&=\frac{V_{\alpha} V_\beta}{4}
\int \frac{dE}{2\pi}{\rm Tr} \left[{\bf F}_0(E)
{\bf F}_1(E)[{\bf F}_2(E_+) e^{i\Delta_{\beta \alpha}} \right. 
\nonumber \\
&&\left.+e^{-i\Delta_{\beta \alpha}} {\bf F}_2(E_-)] 
{\bf F}_3(E) \right] 
\label{show}
\end{eqnarray}
where
\begin{equation}
{\bf G} \equiv {\bf F}_1 {\bf V}_{\alpha} {\bf F}_2 {\bf V}_{\beta}
{\bf F}_3
\end{equation}
and $F_i$'s ($i=0,1,2,3$) satisfy $F_i(t_1,t_2)=F_i(t_1-t_2)$.

Taking the Fourier transform, 
\begin{equation}
{\bf F}(t)=\frac{1}{2\pi }\int dE  e^{-iE t}{\bf F}(E)
\end{equation}
we obtain
\begin{eqnarray}
&&B=\frac{1}{\tau}\int_0^\tau dt \int_{-\infty }^{t}dt_{1}\int 
\frac{dE}{2\pi }{\bf F}_0(E) e^{-iE(t_{1}-t)}
\int dxdy \nonumber \\
&&[{\bf F}_1(t-x)V_{\alpha}(x) {\bf F}_2(x-y) V_{\beta}(y)
{\bf F}_3(y-t_{1})] 
\nonumber \\
&&=\frac{1}{\tau}\int_0^\tau dt \int \prod_{i=1,5} 
\frac{dE_{i}}{2\pi} \int \frac{dE}{2\pi}
\int_{-\infty }^{t}dt_{1}\int dx dy \nonumber \\
&&{\bf F}_0(E){\bf F}_1(E_1) {\bf V}_\alpha (E_{2})
{\bf F}_2(E_3) {\bf V}_\beta (E_4) {\bf F}_3(E_{5})  \nonumber \\
&&e^{i(E-E_1)t}e^{i(E_5-E)t_1}e^{i(E_1-E_2-E_3)x}e^{i(E_3-E_4-E_5)y}
\nonumber
\end{eqnarray}
where 
\begin{equation}
{\bf V}_{\alpha}(E)=\pi V_{\alpha}[e^{i\phi_{\alpha}}\delta (E_+)
+e^{-i\phi_{\alpha}}\delta (E_-)]
\label{fourier}
\end{equation}

Integrating over x and y yields,
\begin{eqnarray}
&&B=\frac{1}{\tau}\int_0^\tau dt \int \prod_{i=1,5} \frac{dE_{i}}{2\pi} 
\int \frac{dE}{2\pi} \int_{-\infty }^{t}dt_{1}{\bf F}_0(E) 
\nonumber \\
&& {\bf F}_1(E_{1}) {\bf V}_\alpha (E_2) {\bf F}_2(E_{3})
{\bf V}_\beta (E_4) {\bf F}_3(E_{5}) e^{i(E-E_{1})t}
\nonumber \\
&& e^{i(E_{5}-E)t_{1}}(2\pi )^{2}  
\delta (E_{1}-E_{2}-E_{3})\delta (E_{5}-E_{4}-E_{5})  \nonumber
\end{eqnarray}

Integrating over $t_1,t$ and using Eq.(\ref{fourier}), we have 
\begin{eqnarray}
&&B=\int \frac{dE_{2}}{2\pi } \frac{dE_{4}}{2\pi} \frac{dE_{5}}{2\pi}
\int \frac{dE}{2\pi }\frac{{\bf F}_0(E)}{i[E_5-E-i\delta]} 
\nonumber \\
&& {\bf F}_1(E_{2}+E_{4}+E_{5}) {\bf V}_{\alpha}(E_{2})
{\bf F}_2(E_{4}+E_{5}) \nonumber \\
&& {\bf V}_{\beta}(E_{4}) {\bf F}_3(E_5)
\delta(E_2+E_4)
\nonumber \\
&&=\frac{V_\alpha V_\beta}{4}\int \frac{dE_{5}}{2\pi } \frac{dE}{2\pi }
\frac{{\bf F}_0(E)}{i[E_{5}-E-i\delta ]}
{\bf F}_1(E_{5}) \nonumber \\
&& [{\bf F}_2(E_{5}+\omega) e^{i\Delta_{\beta
\alpha}}+e^{-i\Delta_{\beta \alpha}} {\bf F}_2(E_{5}-\omega )]
{\bf F}_3(E_{5})  \nonumber 
\end{eqnarray}
Using the theorem of residue, we have 
\begin{equation}
\int dE \frac{F_0(E)}{i[E_5-E-i\delta]}=2\pi F_0(E_5)
\end{equation}
we thus obtain Eq.(\ref{show}).

\bigskip
\bigskip
\bigskip
\noindent{$^{a)}$ Electronic mail: jianwang@hkusub.hku.hk}

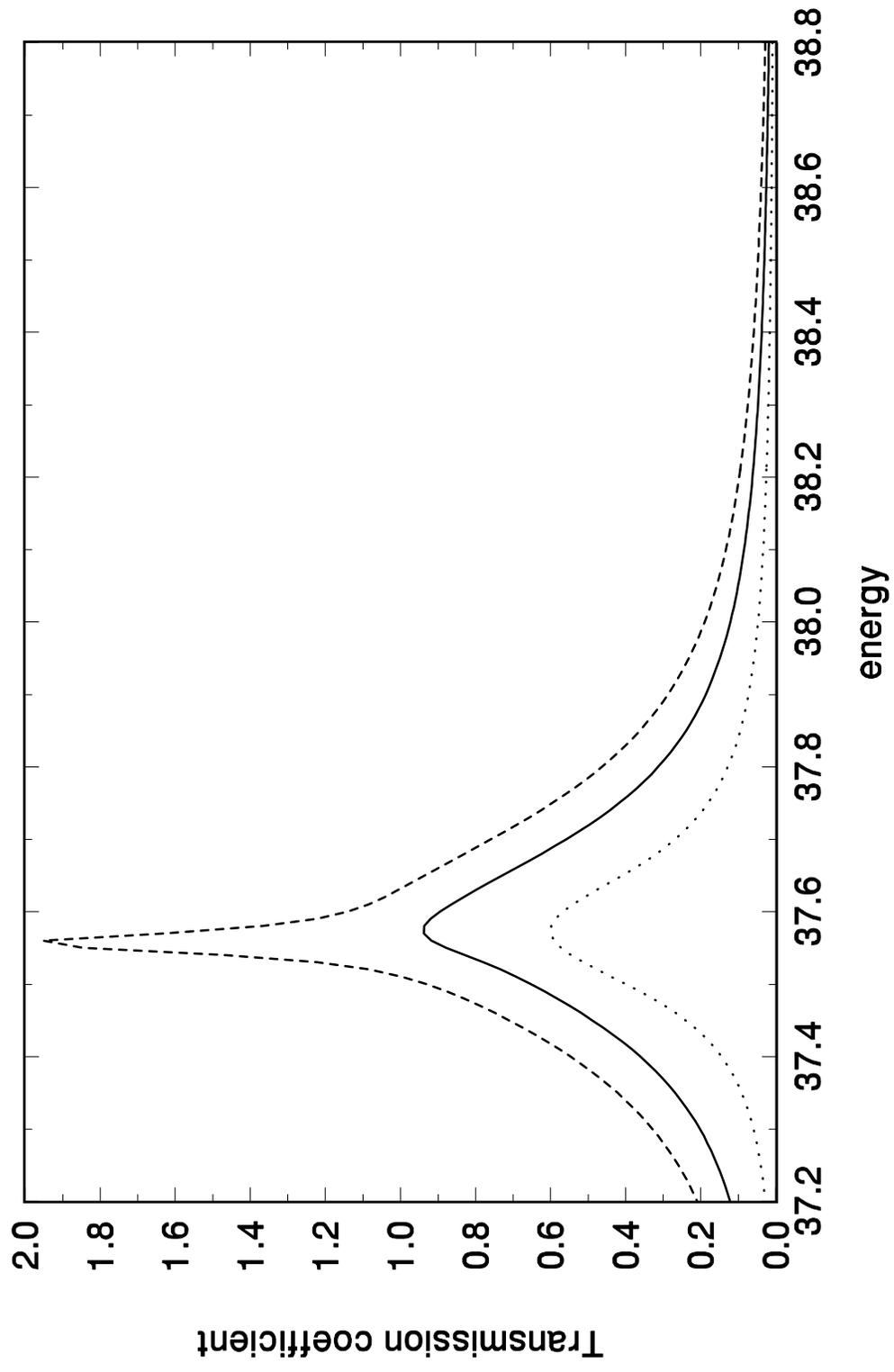
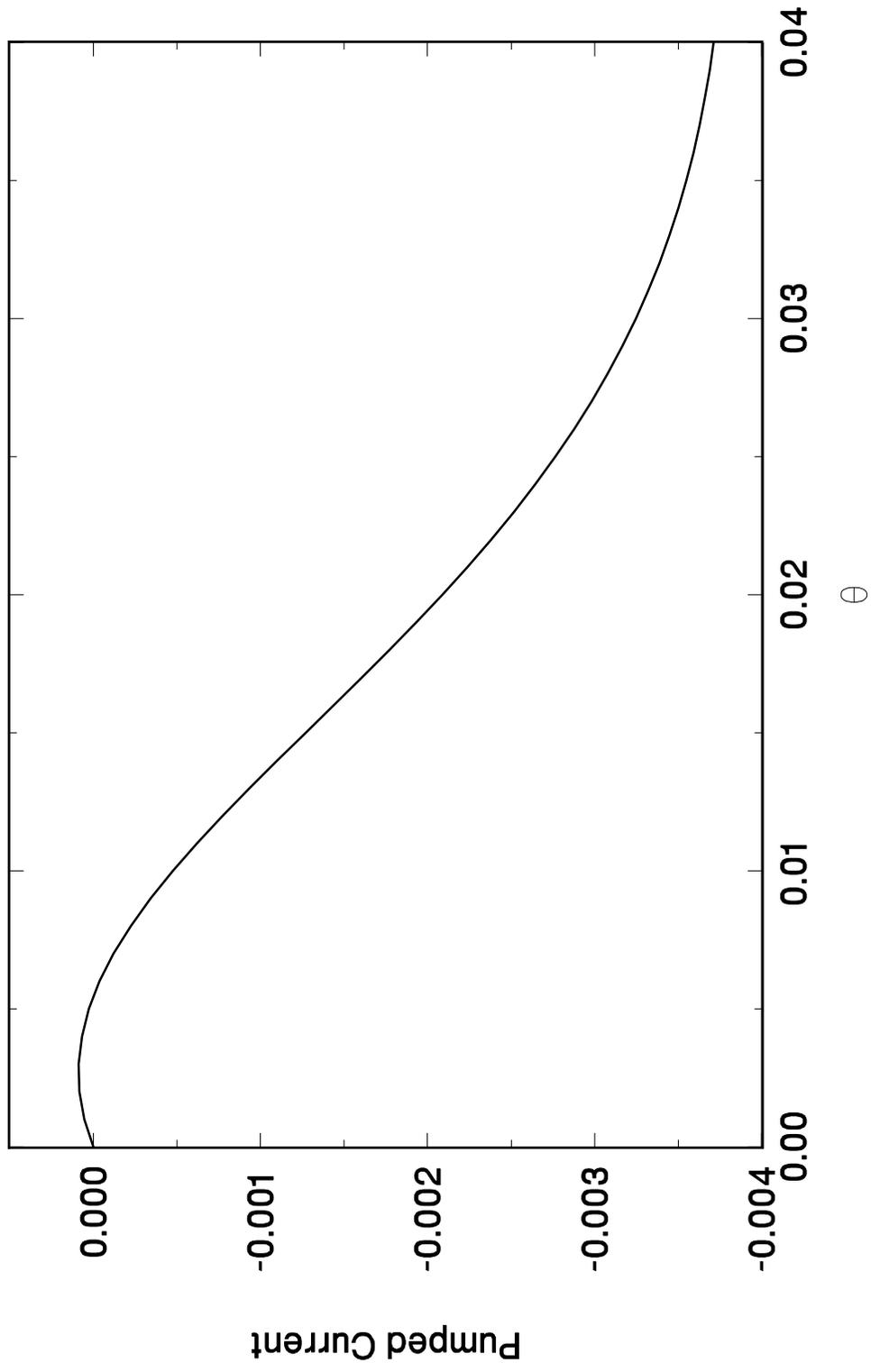
\begin{figure}
\caption{
The transmission coefficient as a function of Fermi energy at different 
angle $\theta$ between the magnetizations of two leads:  $\theta=0$
(dashed line), $\theta=\pi/2$ (solid line), and $\theta=\pi$ (dotted line).
}
\end{figure}

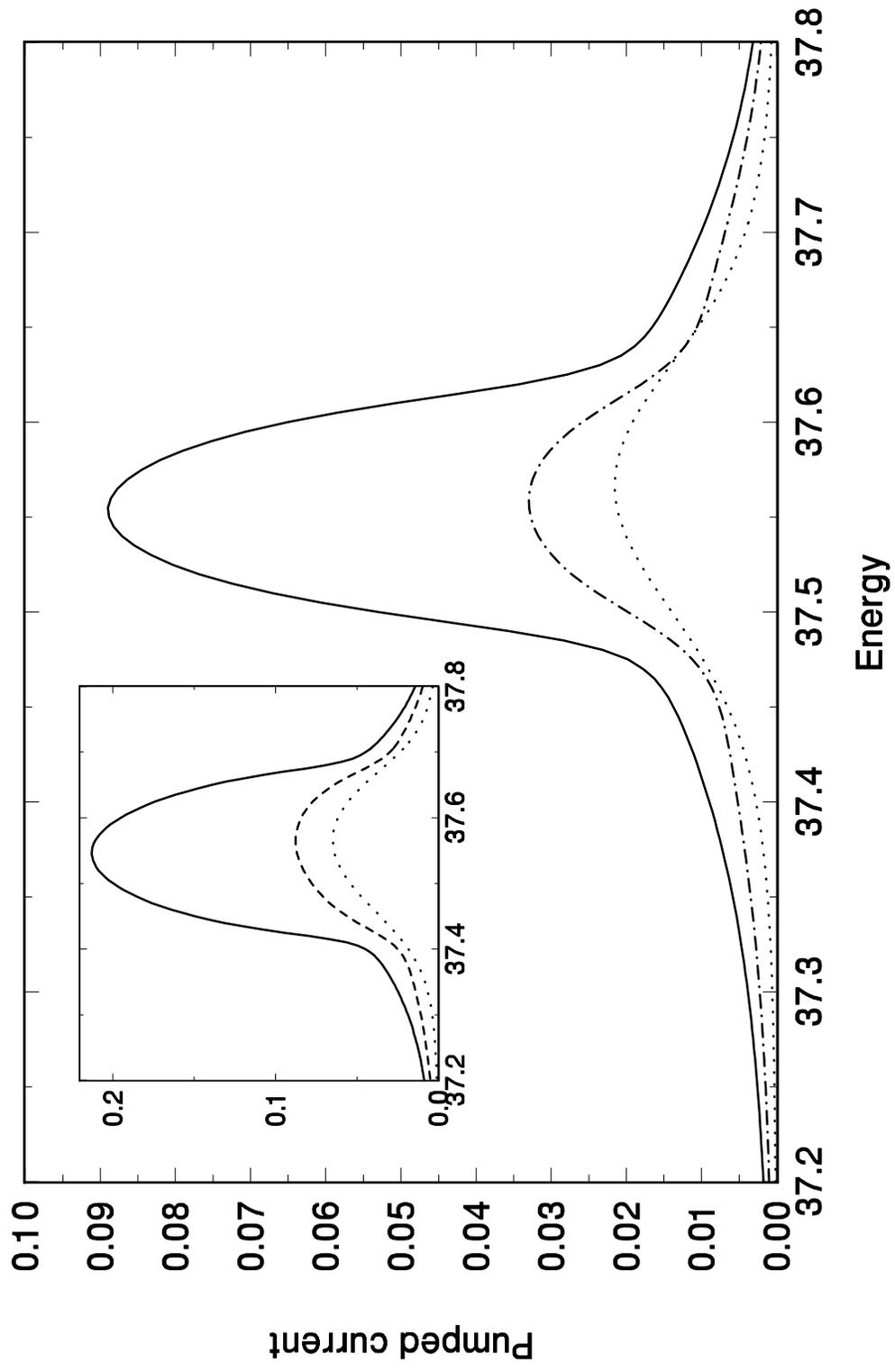
\begin{figure}
\caption{
The pumped current as a function of Fermi energy at different 
$\theta$:  $\theta=0$ (solid line), $\theta=\pi/2$ (dot-dashed line), 
and $\theta=\pi$ (dotted line). Other parameters are $\phi=\pi/2$ and 
$V_p=0.05V_0$. Inset: the same as the main figure except $V_p=0.1 V_0$.
}
\end{figure}

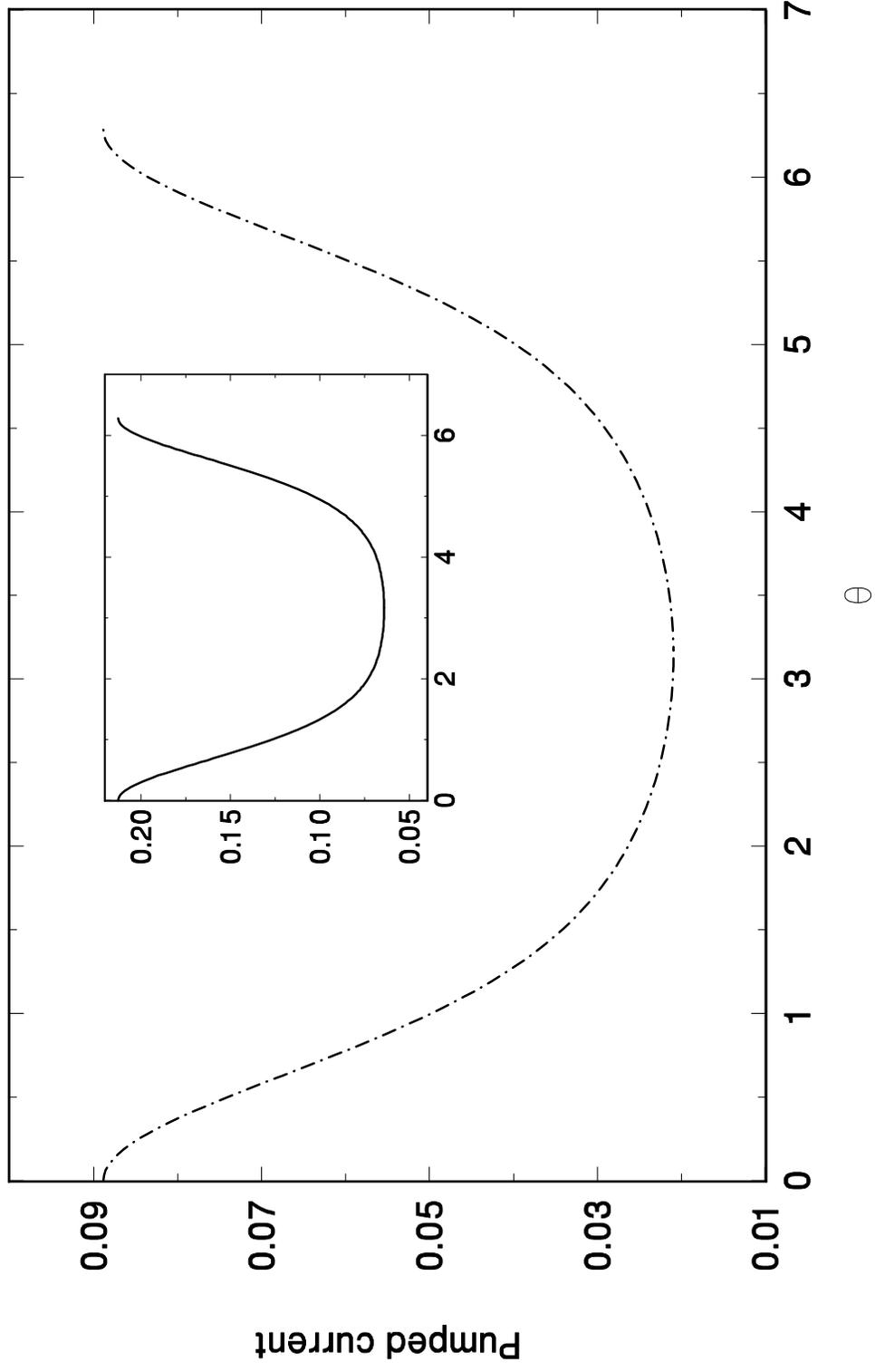
\begin{figure}
\caption{
The pumped current as a function of $\theta$. Here $\phi=\pi/2$, 
$E_F=37.55$ and $V_p=0.05 V_0$. Inset: the same as main figure except
$V_p=0.1 V_0$. 
}
\end{figure}

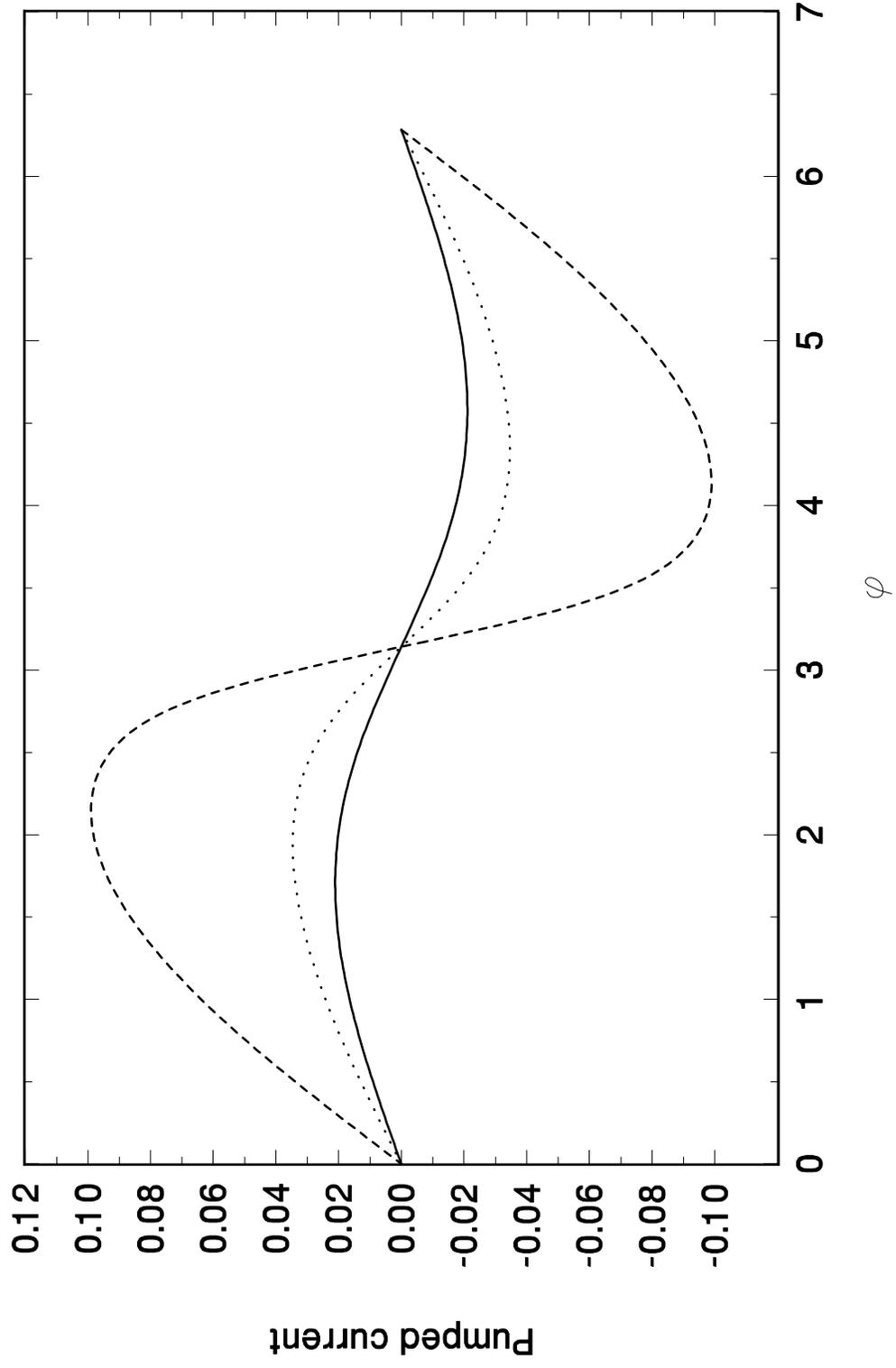
\begin{figure}
\caption{
The pumped current as a function of phase difference $\phi$ at different
$\theta$. $\theta=0$ (dashed line), $\theta=\pi/2$ (dotted line), 
and $\theta=\pi$ (solid line). Other parameters are $E_F=37.55$ and 
$V_p=0.05 V_0$. 
}
\end{figure}

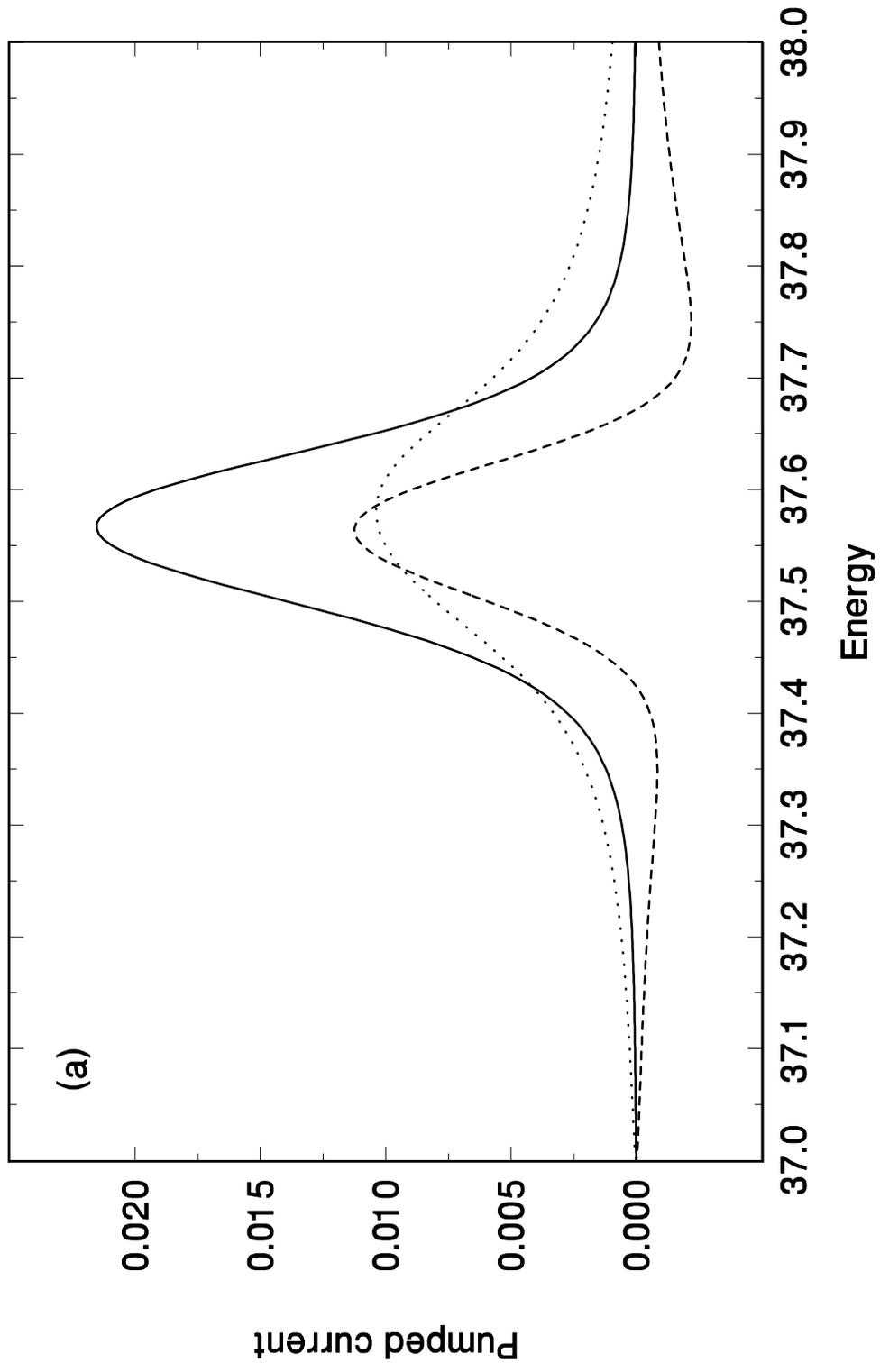
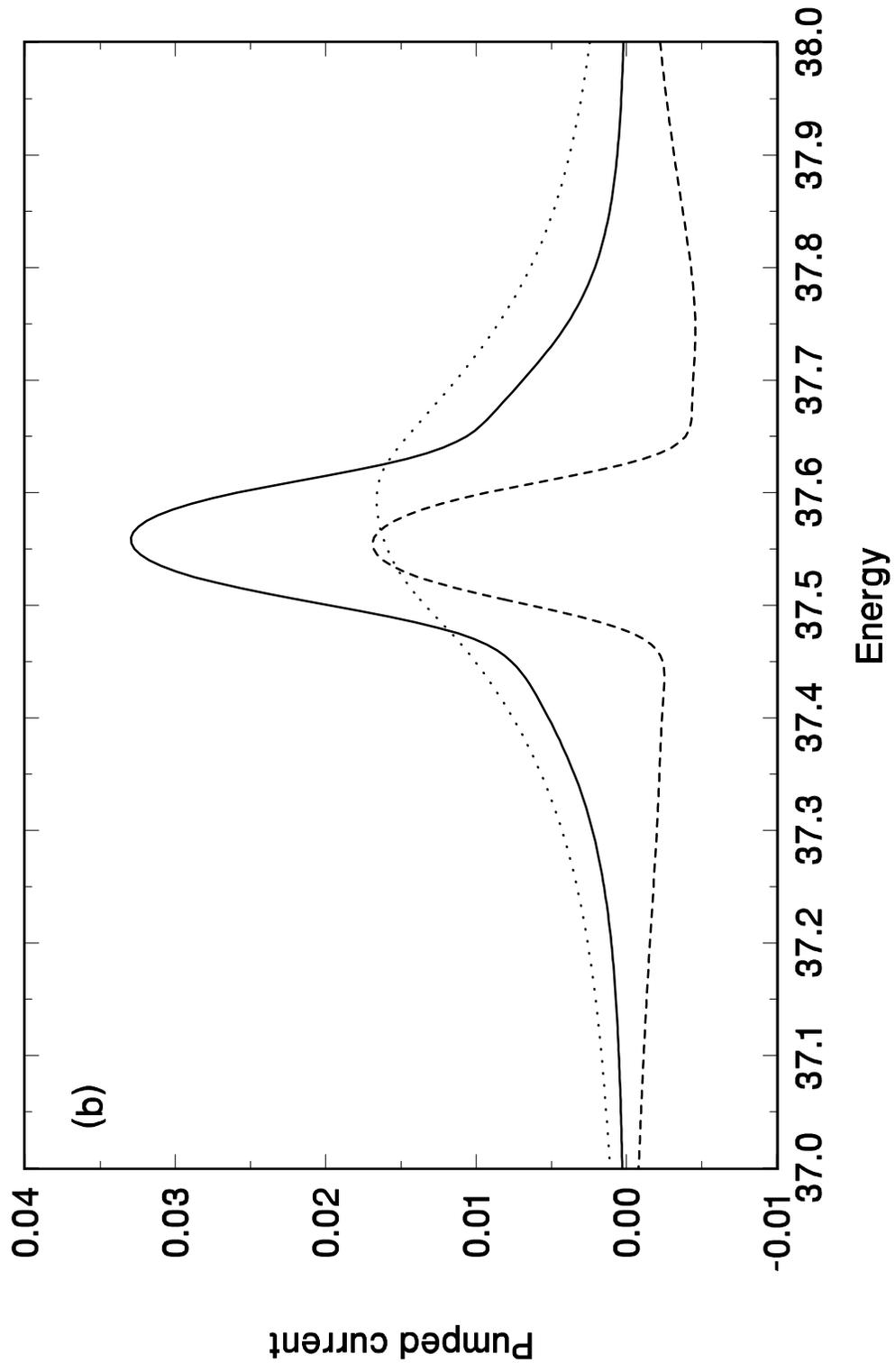
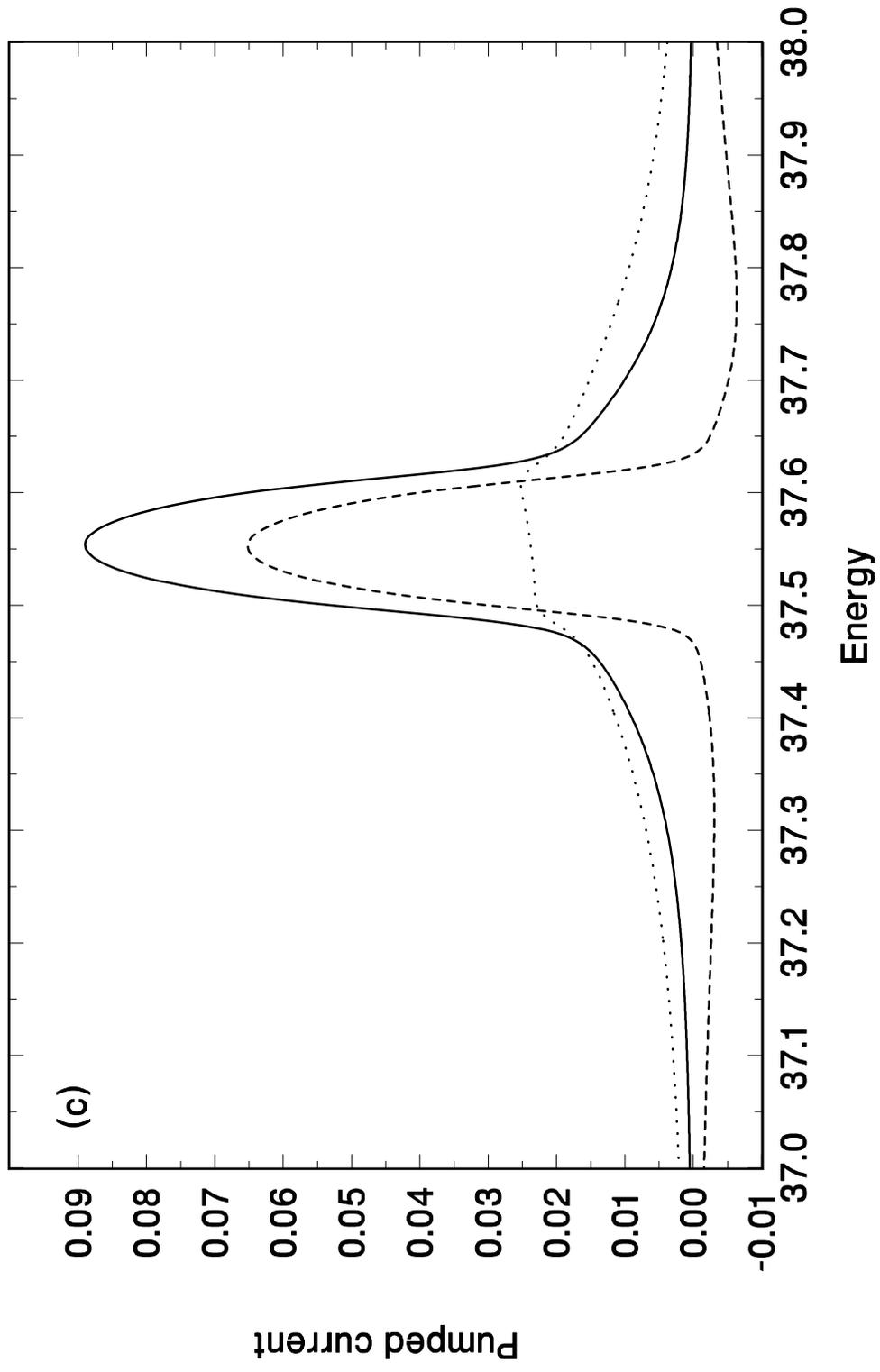
\begin{figure}
\caption{
The pumped current as a function of Fermi energy with an external bias 
$V_L-V_R=-0.02\omega$ at different $\theta$: (a). $\theta=\pi$. 
(b). $\theta=\pi/2$. (c). $\theta=0$. 
In (a), (b) and (c), solid line: pumped current, 
dotted line: current due to external bias, dashed line: total current. 
Here $\phi=\pi/2$ and $V_p=0.05V_0$.
}
\end{figure}

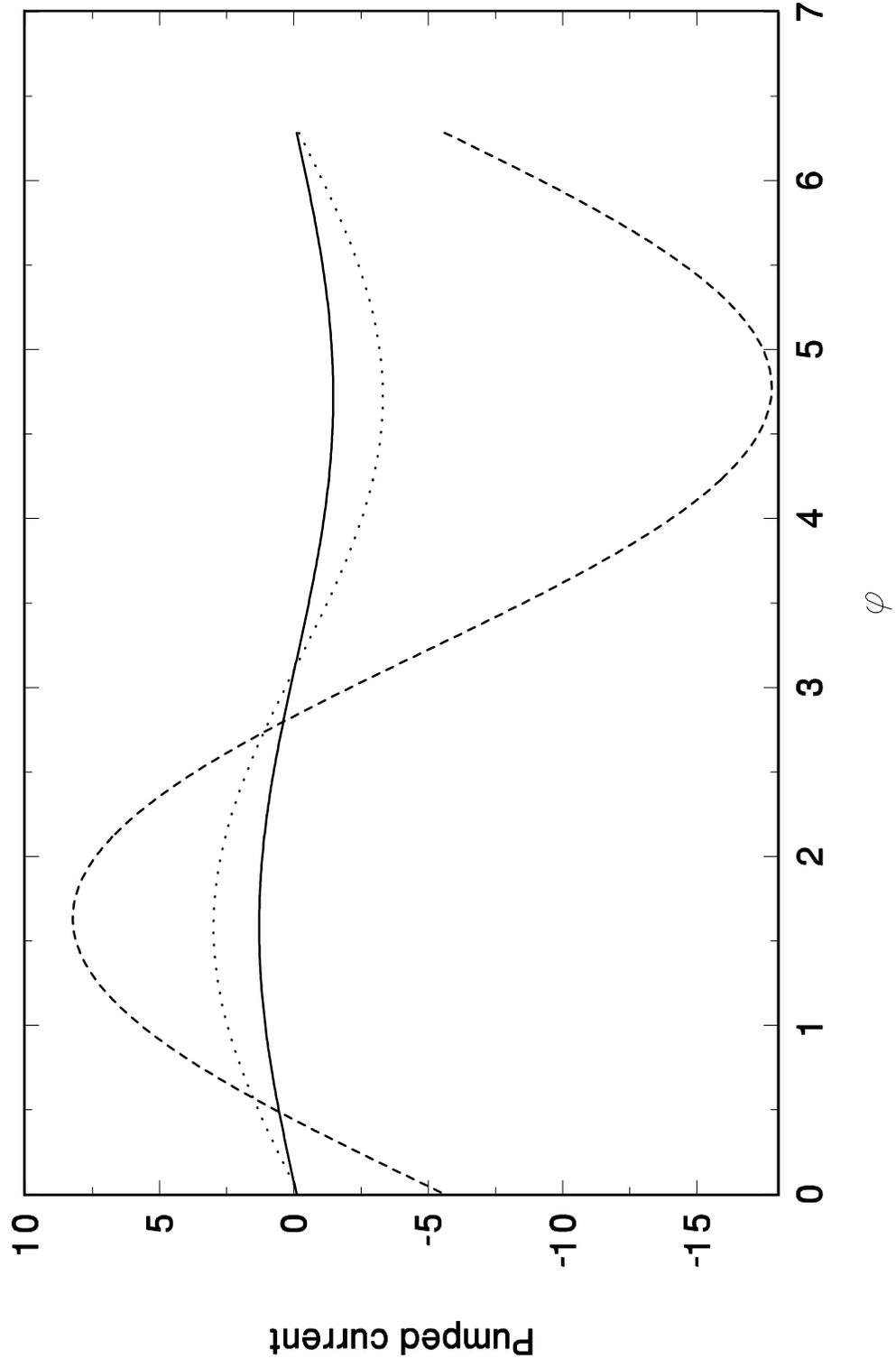
\begin{figure}
\caption{
The pumped current as a function of $\phi$ at finite frequency. 
Dashed line: $\theta=0$, dotted line: $\theta=\pi/2$, solid line: 
$\theta=\pi$.  Here $E_F=37.55$ and $\omega=0.002$. 
}
\end{figure}

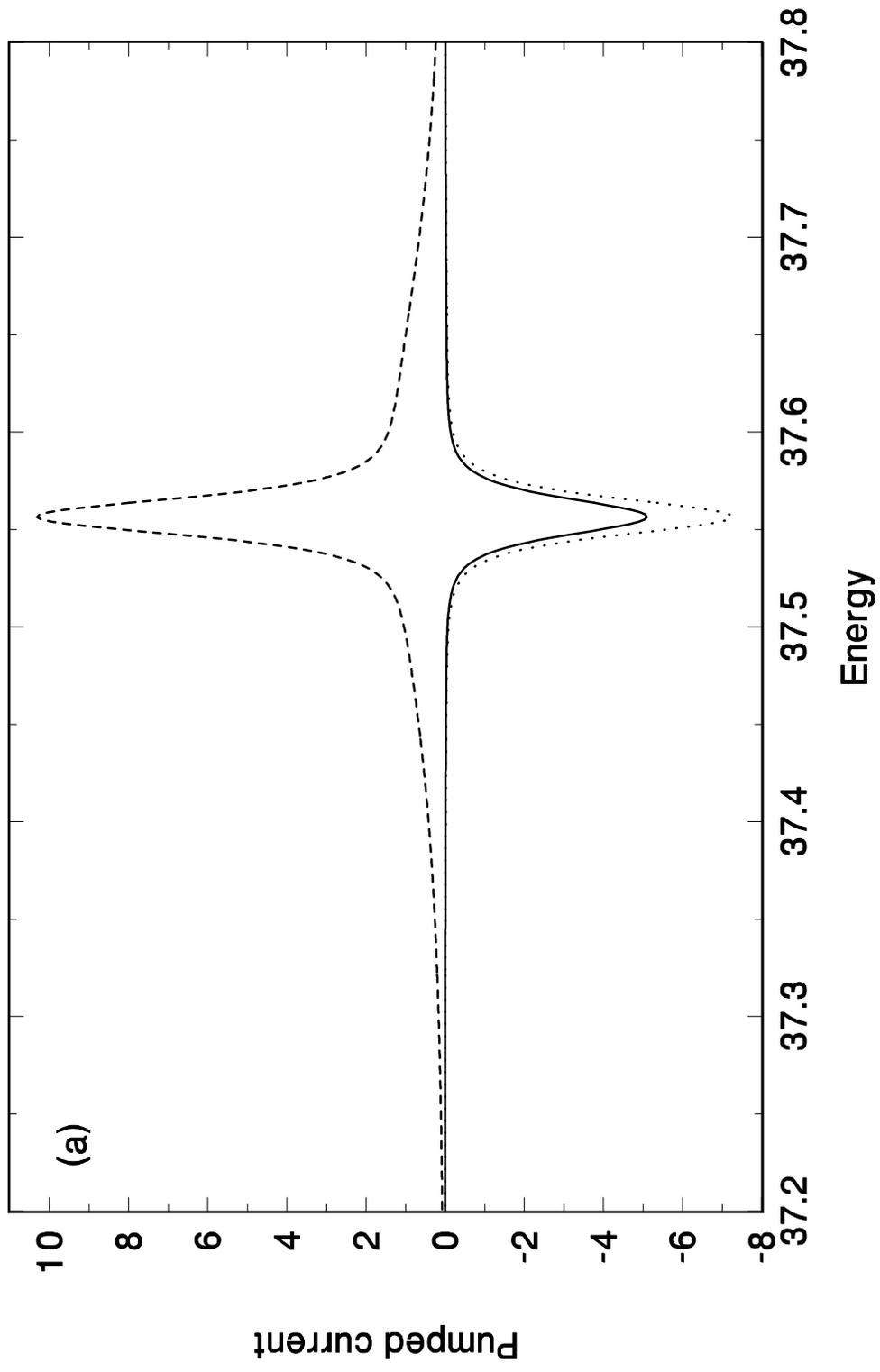
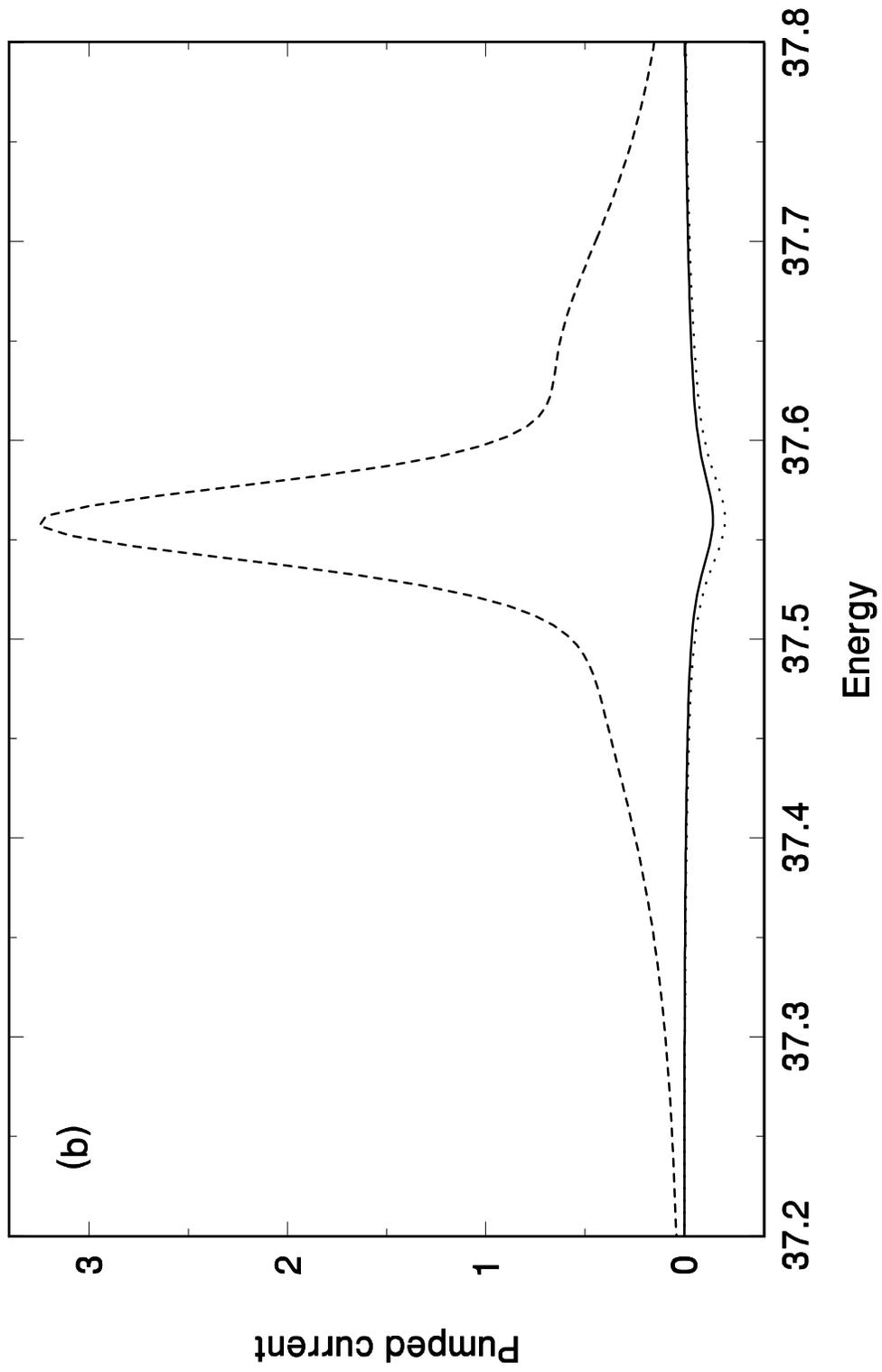
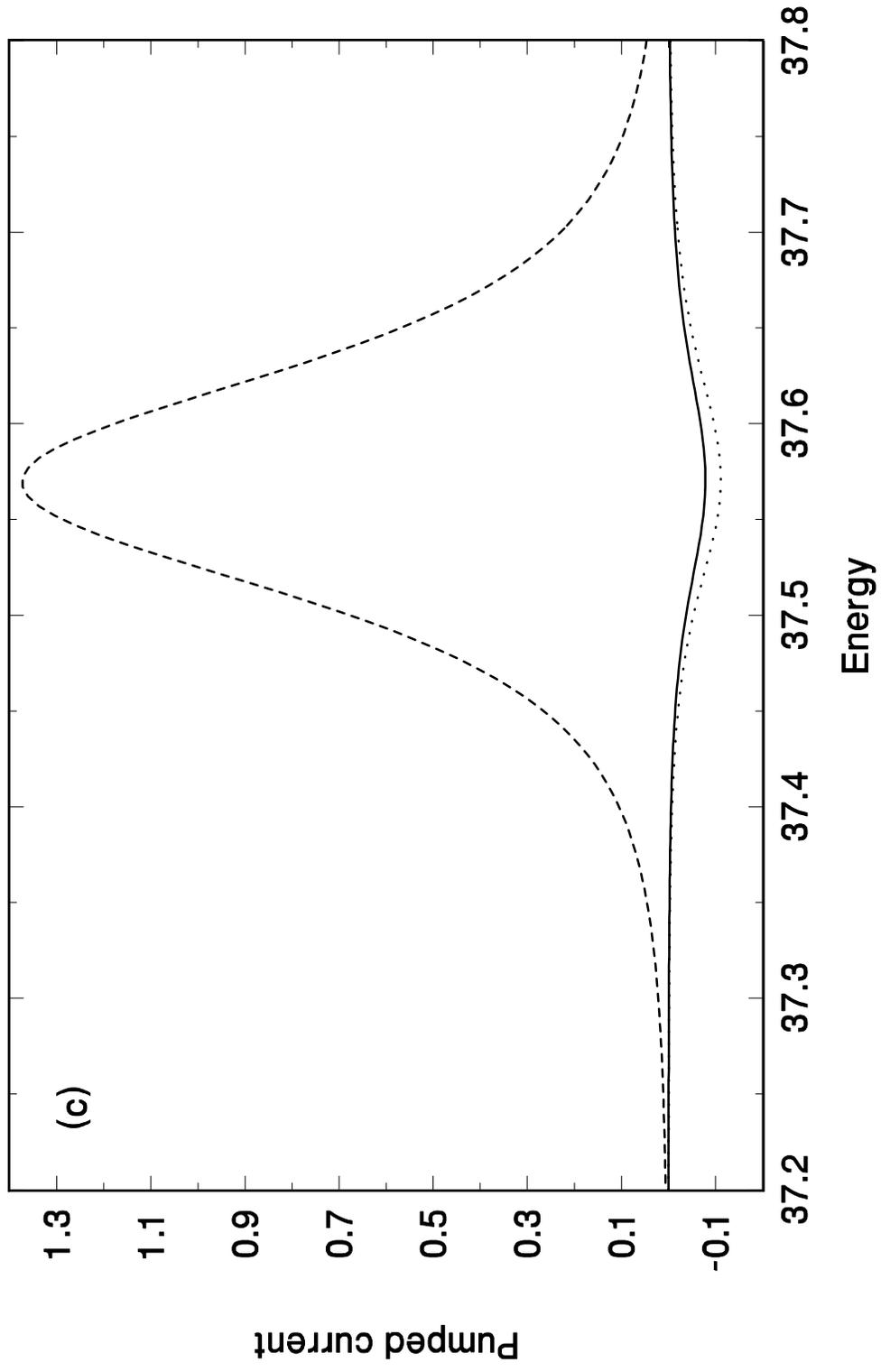
\begin{figure}
\caption{
The pumped current as a function of Fermi energy at finite frequency at
different $\theta$: (a). $\theta=0$. (b). $\theta=\pi/2$. (c). 
$\theta=\pi$. In (a), (b) and (c),  dotted line: $\phi=0$, dashed line: 
$\phi=\pi/2$, solid line: $\phi=\pi$.  Here $\omega=0.002$. 
}
\end{figure}

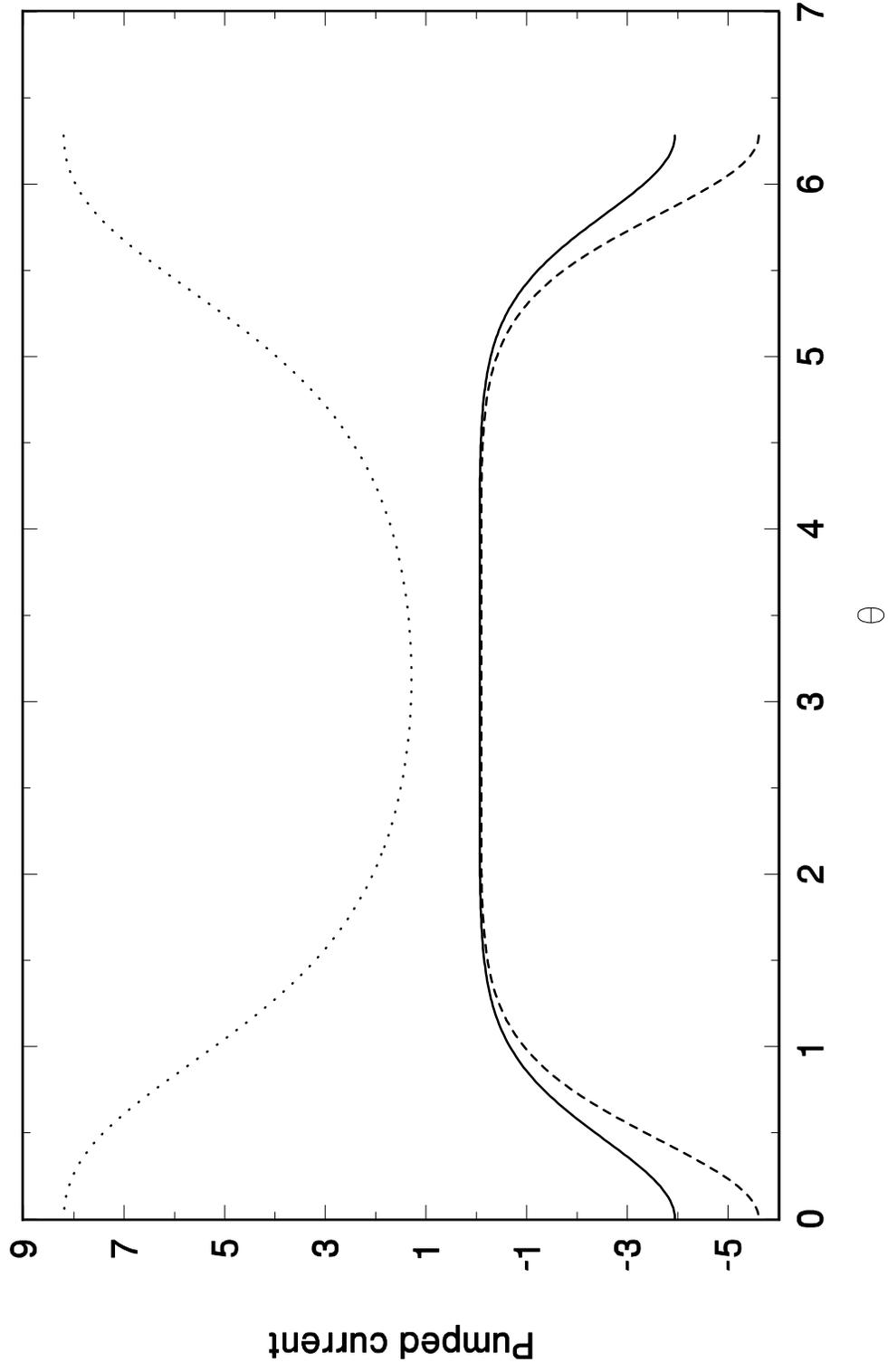
\begin{figure}
\caption{
The pumped current as a function of $\theta$ at finite frequency. 
Here dashed line: $\phi=0$, dotted line: $\phi=\pi/2$, solid line: 
$\phi=\pi$. Other parameters are $E_F=37.55$ and $\omega=0.002$.
}
\end{figure}

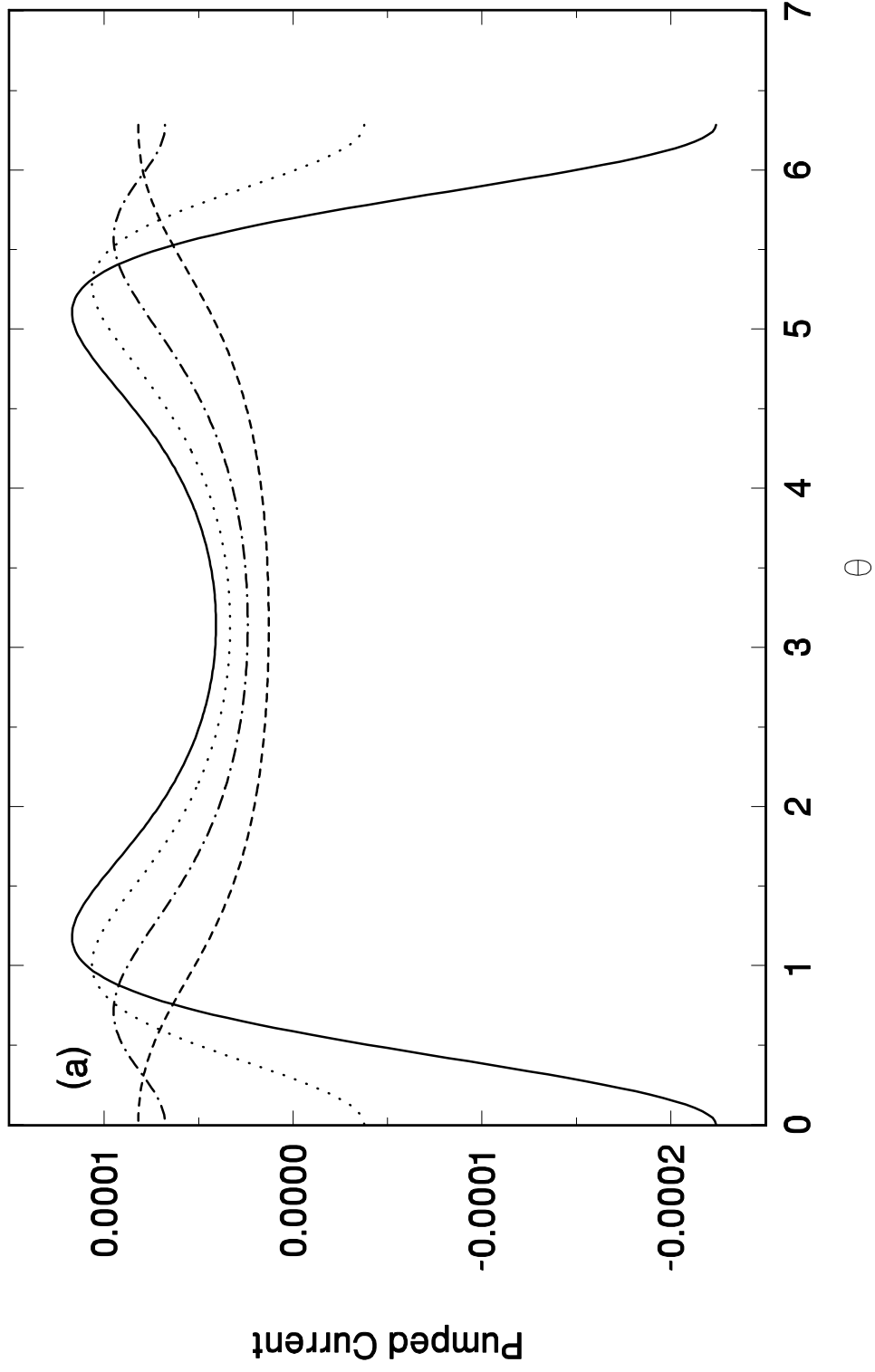
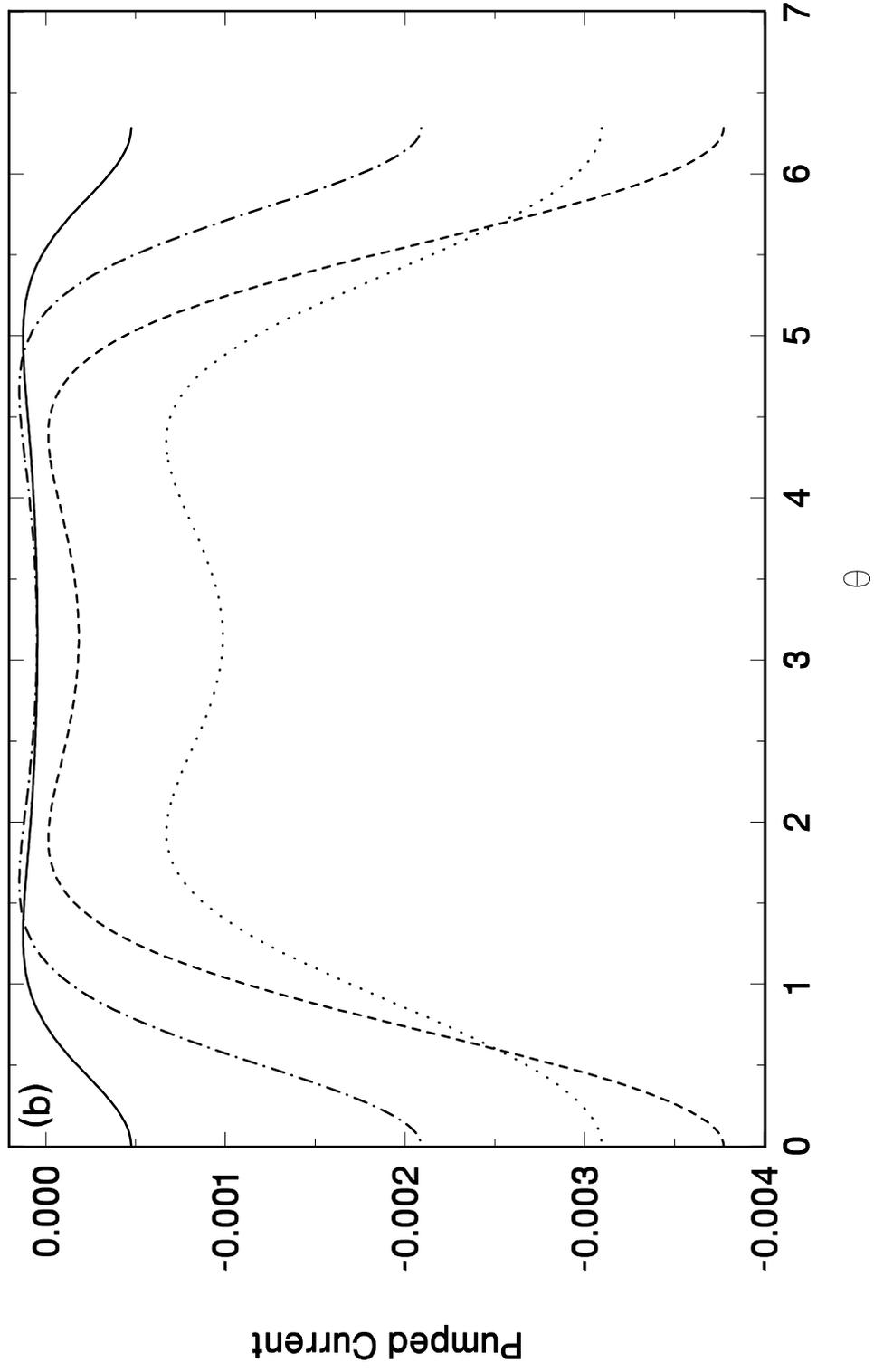
\begin{figure}
\caption{
The pumped current as a function of $\theta$ at different frequencies.
(a). $\omega=0.002$ (short dashed line), $0.004$ (dot-dashed line), $0.006$
(dotted line), $0.008$ (solid line). (b). $\omega=0.01$ (solid line), 
$0.02$ (dot-dashed line), $0.05$ (short dashed line), $0.1$ (dotted
line).  Other parameters are $E_F=37.55$ and $\phi=\pi/2$.
}
\end{figure}

\begin{figure}
\caption{
The pumped current as a function of frequency. Here $\theta=0$, 
$\phi=\pi/2$ and $E_F=37.55$.
}
\end{figure}


\begin{thebibliography}{00}
\bibitem{foot2}
For the system of Fe/Ge/Fe with $a=100A$, the energy uint is $E=4.64 meV$
which corresponds to frequency $\omega=1.1 \times 10^{12}$ Hz. The
unit of the pumped current in the adiabatic regime depends on the
pumping frequency $\omega$. If $\omega=100$MHz, then the pumped current
is $1.6 \times 10^{-11}$A. For pumped current at finite frequency, the 
unit of current is $2\times10^{-6}$A.
\end{thebibliography}
\end{document}